\documentclass[twocolumn]{jpsj3} 
\def\etal{{\it et\ al.}}

\newcommand{\lsim}
 {\ \raise.35ex\hbox{$<$}\kern-0.75em\lower.5ex\hbox{$\sim$}\ }
\newcommand{\gsim}
 {\ \raise.35ex\hbox{$>$}\kern-0.75em\lower.5ex\hbox{$\sim$}\ }
\def\journal #1#2#3#4{#1 {\bf #2} (#4) #3}
\def\PR{Phys.\ Rev.}
\def\PRA{Phys.\ Rev.\ A}
\def\PRB{Phys.\ Rev.\ B}
\def\PRL{Phys.\ Rev.\ Lett.}

\def\IJMP{Int.\ J.\ Mod.\ Phys.}

\def\JLTP{J.~Low Temp.~Phys.}

\def\JPCS{J.\ Phys.\ Chem.\ Solids}

\def\JPSJ{J.\ Phys.\ Soc.\ Jpn.}

\def\NP{Nat.~Phys.}

\def\RMP{Rev.\ Mod.\ Phys.}
\def\PTP{Prog.\ Theor.\ Phys.}

\def\EPL{Europhys.\ Lett.}

\title{Effect of Doublon-Holon Binding on Mott transition---Variational
Monte Carlo Study of Two-Dimensional Bose Hubbard Models}

\author{Hisatoshi {\sc Yokoyama}\thanks{E-mail address: 
                          yoko@cmpt.phys.tohoku.ac.jp}, 
        Tomoaki {\sc Miyagawa} and 
        Masao {\sc Ogata}$^{1}$
}
\inst{Department of Physics, Tohoku University, 
      Sendai 980-8578 \\
$^{1}$Department of Physics, University of Tokyo, 
      Bunkyo-ku, Tokyo 113-0033
}
\abst{
To understand the mechanism of Mott transitions in case of no magnetic 
influence, superfluid-insulator (Mott) transitions in the $S=0$ Bose 
Hubbard model at unit filling are studied on the square and triangular 
lattices, using a variational Monte Carlo method. 
In trial many-body wave functions, we introduce various types of 
attractive correlation factors between a doubly-occupied site (doublon, D) 
and an empty site (holon, H), which play a central role for Mott 
transitions, in addition to the onsite repulsive (Gutzwiller) factor. 
By optimizing distance-dependent parameters, we study various properties 
of this type of wave functions. 
With a hint from the Mott transition arising in a completely D-H 
bound state, we propose an improved picture of Mott transitions, 
by introducing two characteristic length scales, the D-H binding length 
$\xi_{\rm dh}$ and the minimum D-D exclusion length $\xi_{\rm dd}$.
Generally, a Mott transition occurs when $\xi_{\rm dh}$ becomes comparable 
to $\xi_{\rm dd}$. 
In the conductive (superfluid) state, domains of D-H pairs overlap 
with each other ($\xi_{\rm dh}>\xi_{\rm dd}$); thereby D and H can 
propagate independently as density carriers by successively exchanging 
the partners.
In contrast, intersite repulsive Jastrow (D-D and H-H) factors have 
little importance for the Mott transition. 
} 

\kword{Mott transition, Bose Hubbard model, two dimensions, 
superfluid, insulator, variational Monte Carlo Method, 
doublon-holon binding}

\begin{document}
\maketitle
\section{Introduction\label{sec:intro}}
After early studies of superfluid-insulator or Mott transitions 
in interacting Bose systems \cite{Fisher}, an experimental examples 
have been realized using ultracold dilute gases of bosonic atoms 
in various optical lattices.\cite{Greiner,1D,Kohl,Spielman} 
The essence of these systems is captured by a spinless Bose Hubbard model 
with a harmonic confinement potential.\cite{Jaksch,Bloch-Rev} 
Aside from this one-body potential, this basic model is important 
on the theoretical ground that a Mott transition can be studied 
without cares of magnetic influence, unlike typical fermionic cases. 
Most researchers are certain that, for the Fermi Hubbard models, 
metal-insulator transitions take place at infinitesimal correlation 
strength on the hypercubic lattice in any dimension.\cite{Lieb-Wu}  
Such transitions at $U/t=0$ ($U$: onsite-interaction strength, $t$: 
hopping integral) require elements other than competition between 
itineracy and bare interaction of particles, like magnetic correlation. 
In contrast, spinless Bose Hubbard models bring about Mott transitions 
at moderate finite values of $U/t$.\cite{critical,Uc-1d,Uc-3d1,Uc-3d2} 
\par

To date, Mott transitions in the $S=0$ Bose Hubbard model have been 
studied with various methods. 
For the square lattice of our interest, properties of $T_{\rm c}$, 
superfluid density, etc.~were studied, applying a quantum Monte Carlo 
(QMC) method earlier to small systems\cite{QMC-Krauth} and later 
to larger systems;\cite{QMC-Wessel,QMC-Sansone}
a ground-state phase diagram in the plane of chemical potential 
and interaction strength was constructed using a strong-coupling 
expansion \cite{Monien}. 
These studies estimated the critical interaction strength of 
Mott transitions at $U_{\rm c}/t=16.4$-$16.7$ for the particle 
density of unit filling, $n=1$ ($n=N/N_{\rm s}$ with $N$: 
particle number, $N_{\rm s}$: site number) at $T=0$. 
Thus, the existence of a Mott transition has been embodied, but the 
mechanism of the transition is still not clear. 
\par

Variational Monte Carlo (VMC) approaches\cite{McMillan,YS1,Umrigar} are 
very useful to analyze the physics of Mott transitions, because one 
can directly and exactly treat wave functions for any values of $U/t$. 
As variational approaches to the Bose Hubbard model, wave functions 
with only onsite correlation factors, which correspond to the celebrated 
Gutzwiller wave function\cite{Gutz} (GWF, $\Psi_{\rm G}$) for fermions, 
were studied first \cite{BHM-GWF1,BHM-GWF2}. 
In contrast to for fermions, GWF for bosons is solved analytically 
without additional mean-field-type approximations \cite{GA} in 
arbitrary dimensions, and yield a Brinkman-Rice-type (BR) 
superfluid-insulator transition \cite{BR} at finite $U$ 
($\equiv U_{\rm BR}$, see Fig.~\ref{fig:EvsU-mu}). 
In the insulating side of $U_{\rm BR}/t$, however, all the lattice sites 
are occupied with exactly one particle and the hopping completely ceases, 
namely, $\Psi_{\rm G}\rightarrow\prod_{j=1}^Nb_j^\dag|0\rangle$ and 
the total energy vanishes ($E=0$). 
This result, which apparently contradicts experiments and reliable 
theories like the strong-coupling expansion, is caused by an oversimplified 
setup of the wave function, in which the effect of density fluctuation 
should be included. 
To remedy this drawback, it is crucial to add appropriate intersite 
correlation factors to GWF. 
In this line, recent VMC studies\cite{Capello} emphasized the importance 
of an ordinary type of long-range Jastrow factor for the transition. 
\par

In this paper, we first show that an attractive correlation factor 
between a doubly-occupied site (doublon: D) and an empty site 
(holon: H) plays a leading role to induce the Mott transition 
for the present bosonic model. 
In fact, D-H near-neighbor correlations have long been studied 
for Fermi Hubbard models,\cite{Castellani,Kaplan,Fazekas,YS} 
and the present authors have shown that near-neighbor D-H binding 
factors and their analogs are capable of inducing Mott transitions 
for attractive Hubbard models\cite{Y-PTP}, a repulsive Hubbard model 
on an extended square lattice\cite{YTOT,YOT} and on an anisotropic 
triangular lattice.\cite{Watanabe} 
In this mechanism, in the insulating regime, a doublon and a holon 
as density carriers are confined in the range of near-neighbor sites; 
namely, density fluctuation is localized. 
In this work, we extend the D-H binding correlation factor to various 
long-range types in order to corroborate the above conclusion, 
and propose a renewed picture of conduction, which can explain 
a Mott transition arising in a completely D-H bound state. 
Namely, the Mott transition occurs, when a D-H binding length 
$\xi_{\rm dh}$ regulated by the D-H attractive correlation becomes 
comparable to the minimum D-D (H-H) distance $\xi_{\rm dd}$. 
This corresponds to the condition that a D-H pair comes to stop 
exchanging the partner with nearby D-H pairs.  
Furthermore, we introduce a repulsive Jastrow factor to check that 
it is of no importance for the Mott transition. 
These properties for bosons are fundamentally common to fermions, 
unless magnetic correlation is explicitly introduced.\cite{Miyagawa} 
\par

This paper is organized as follows: 
In \S\ref{sec:formalism}, the model and the trial wave functions 
used in this paper are introduced. 
In \S\ref{sec:results}, the properties of the Bose Hubbard model 
as to the Mott transition are studied with a short-range D-H binding 
factor. 
In \S\ref{sec:long-range}, we consider the effect of long-range D-H 
attractive correlation factors, and of D-D repulsive factors. 
In \S\ref{sec:picture}, we discuss a renewed picture of the Mott 
transition by introducing two characteristic length scales. 
In \S\ref{sec:sum}, a concise summary is given. 
In Appendix, we briefly note the setup and condition of the VMC 
calculations implemented in this paper. 
\par

Parts of the results in this paper have been published 
before.\cite{SNS2007,YMO}
\par

\section{Formalism\label{sec:formalism}}
After defining the model in \S\ref{sec:model}, we introduce trial wave 
functions with various types of D-H attractive correlation factors 
and a long-range D-D (and H-H) repulsive factor in \S\ref{sec:wf}. 
\par

\subsection{Bose Hubbard Model\label{sec:model}}
We consider the $S=0$ Bose Hubbard model on the square (SQL) and 
triangular (TAL) lattices with only homogeneous nearest-neighbor (NN)
hopping:
\begin{equation}
H = -t \sum_{\langle ij\rangle} {(b_i^\dag b_j + b_j^\dag b_i)} 
+ \frac{U}{2}\sum_j n_j(n_j-1), 
\label{eq:model} 
\end{equation}
where $b_j$ ($n_j=b^\dag_j b_j$) denotes an annihilation (number) 
operator of a boson on the site $j$. 
We assume $t$, $U\ge 0$, and use $t$ as the energy unit. 
In this paper, we disregard a harmonic confinement potential, 
not only because we allow for the comparison with electron systems in 
solids, but because flat confinement potentials will be expected
in forthcoming experiments of cold atoms.\cite{box} 
We use systems of $L\times L$ ($=N_{\rm s}$) sites with the periodic 
boundary conditions in both $x$ and $y$ directions. 
The kinetic part, $H_t$, in eq.~(\ref{eq:model}) is diagonalized as, 
\begin{equation}
H_t=\sum_{\bf k}\varepsilon_{\bf k}b_{\bf k}^\dag b_{\bf k},
\end{equation}
with 
\begin{equation}
\varepsilon_{\bf k}=\left\{
\begin{array}{ll}
-2t(\cos k_x+\cos k_y), & \mbox{SQL} \\ 
-2t\left[\cos k_x+\cos k_y+\cos(k_x+k_y)\right], & \mbox{TAL} 
\end{array}
\right.
\end{equation} 
using a Fourier transformation, 
\begin{equation}
b_{\bf k}=\frac{1}{\sqrt{N_{\rm s}}}\sum_je^{-i{\bf k}\cdot{\bf r}_j} b_j. 
\label{eq:FT}
\end{equation}
For $U/t=0$, since all the particles condense into the lowest-energy 
level ${\bf k}={\bf 0}=(0,0)$, the ground state is 
\begin{equation}
\Phi_0=\frac{1}{\sqrt{N!}}\ b_{{\bf k}={\bf 0}}^{\dag N}\ |0\rangle, 
\end{equation}
with the eigenenergy per site being $E=-4t$ ($-6t$) for SQL (TAL). 
Using eq.(\ref{eq:FT}), the real-space representation of $\Phi_0$ 
is written as, 
\begin{equation}
\Phi_0\propto
      \sum_{\{{\bf r}\}} b_1^\dag b_2^\dag\cdots b_N^\dag|0\rangle, 
\label{eq:Phireal}
\end{equation}
where $\{{\bf r}\}$ indicates the sum of all the particle configurations 
$\{{\bf r}_1,\cdots,{\bf r}_N\}$, permitting duplicate counting 
in combinations, and we abbreviate $b_{{\bf r}_i}^\dag$ to $b_i^\dag$.
Note that, in contrast to fermionic ground states, the coefficient 
(a permanent of $N\times N$) of every configuration in the sum 
of eq.~(\ref{eq:Phireal}) becomes an identical constant, 
because all the elements in the permanent becomes unity. 
Accordingly, VMC calculations are greatly simplified. 
\par

As mentioned in \S\ref{sec:intro}, the ground-state phase diagram 
and some relevant properties of this model including the Mott critical 
value, $U_{\rm c}/t$, have been studied by reliable methods like QMC. 
Here, we shed light on the properties and mechanism as 
to the Mott transition, taking advantage of the VMC method. 
\par

\subsection{Variational wave functions\label{sec:wf}} 
We adopt a many-body variational approach to tackle the Mott physics 
in eq.~(\ref{eq:model}). 
The simplest trial function is a Bose analog of the Gutzwiller wave 
function\cite{Gutz} (GWF), 
$
\Psi_{\rm G}=g^D\Phi_0,
$
where $g$ is a variational parameter controlling the number of 
multiply occupied sites (multiplon: M) and $D$ is onsite correlation 
operator:
\begin{equation}
D=\frac{1}{2}\sum_jn_j(n_j-1). 
\label{eq:D}
\end{equation}
The correct energy expectation value by $\Psi_{\rm G}$ is given 
by an analog of the Gutzwiller approximation formula,\cite{BHM-GWF1,BHM-GWF2} 
and a Brinkman-Rice-type transition occurs at 
$U_{\rm BR}/t=(1+\sqrt{2})^2z$, where $z$ is the number of 
the NN sites: $U_{\rm BR}/t=23.31$ for SQL, and $34.97$ for TAL. 
However, the description of the insulating state by $\Psi_{\rm G}$ 
is incorrect in that every site is occupied by a single particle and 
density fluctuation is completely suppressed, 
$
\langle H_t\rangle=\langle H_U\rangle=0\  (H_U=UD), 
$ 
in the same way as the Brinkman-Rice transition in the fermionic 
cases.\cite{BR}
\par

At unit filling, a multiplon and a holon are regarded as positive and 
negative particle-density ``carriers" respectively in the background 
of singly-occupied sites of the neutral or average particle density. 
Thus, conductivity depends on whether the motion of such density 
carriers is free or bound; to describe the Mott transition with higher 
fidelity, it is crucial to add an inter-carrier correlation factor, 
especially, a doublon(multiplon)-holon (D-H) factor, 
${\cal P}_Q$:\cite{YS} 
\begin{equation}
\Psi_{\rm DH}={\cal P}_Q\Psi_{\rm G}. 
\label{eq:DH}
\end{equation}
For bosons, multiple site occupation of $p$ $(\ge 0)$ particles 
is allowed; for $U/t=0$, the distribution of the number of sites 
occupied by $p$ particles, $P(p)$, should be Poissonian. 
However, for large values of $U/t$ like in a Mott critical region 
of our interest, $P(p)$ with $p\ge 3$ almost completely 
vanishes, as will be discussed later. 
Then, a multiplon becomes identical to a doublon, and the particle-hole 
symmetry is restored at $n=1$. 
In this context, we treat multiplons and holons symmetrically 
in ${\cal P}_Q$, and often regard M as D for $U\gsim U_{\rm c}/2$. 
\par

In our previous studies for fermions,\cite{YS,Y-PTP,YTOT,YOT,Watanabe} 
we have used only near-neighbor D-H binding factors [{\bf (1)} below], 
which are sufficient to describe Mott transitions. 
In this work, we introduce long-range types of ${\cal P}_Q$, whose 
necessity has long been recognized in an exact-diagonalization study 
in one dimension,\cite{YS} to study the effect of D-H factor more 
in detail. 
We itemize ${\cal P}_Q$ used in this paper below. 
\par 

\begin{figure}[hob]
\begin{center}
\includegraphics[width=8.5cm,clip]{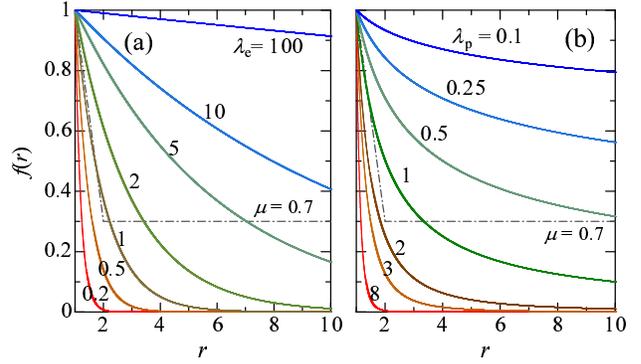} 
\end{center}
\vskip -5mm
\caption{
(Color online) Weight of long-range doublon(D)-holon(H) correlation 
factor in {\bf (2)} as a function of distance between D and 
the nearest H for several values of the parameter $\lambda$, 
eq.(\ref{eq:f(r)}); 
(a) exponentially decaying type in $\Psi_{\rm DH}^{\rm exp}$, and 
(b) power-law decaying type in $\Psi_{\rm DH}^{\rm pwr}$.
For comparison, $f(r)$ of the nearest-neighbor factor 
${\cal P}_Q(0.7,0)$ [eq.(\ref{eq:pq-SQL})] in $\Psi_{\rm DH}^\mu$ 
is plotted with dash-dotted lines. 
}
\label{fig:dh-exp-pw}
\end{figure}

{\bf (1)} 
We extend the above near-neighbor D-H binding factors 
for fermions\cite{errata} to the Bose Hubbard model. 
Following the previous papers, we call this short-range D-H binding 
wave function QWF or $\Psi_{\rm DH}^{\rm \mu(\mu')}$. 
For SQL, we consider up to the second (diagonal)-neighbor correlation: 
\begin{equation} 
{\cal P}_Q={\cal P}_Q(\mu,\mu')
                =(1-\mu)^{\hat{Q}}(1-\mu')^{\hat{Q}'}
\label{eq:pq-SQL}
\end{equation}
where primes indicate diagonal neighbors and,
\begin{equation}
\hat{Q}^{(')}=\sum_i\left[\tilde d_i\prod_{\tau^{(')}} (1-h_{i+\tau^{(')}})
                +h_i\prod_{\tau^{(')}}(1-\tilde d_{i+\tau^{(')}})\right]. 
\label{eq:Q}
\end{equation}
Here, $\tilde d_i$ and $h_i$ are projection operators of multiplon (M) and 
holon (H) on the site $i$, respectively:
\begin{eqnarray}
\tilde d_i|i\rangle &=& \left\{
\begin{array}{ll}
1\ |i\rangle, & \quad\mbox{$|i\rangle$: M} \\
0\ |i\rangle, & \quad\mbox{$|i\rangle$: otherwise}
\end{array}\right., 
                   \\
h_i|i\rangle &=& \left\{
\begin{array}{ll}
1\ |i\rangle, & \quad\mbox{$|i\rangle$: H} \\
0\ |i\rangle, & \quad\mbox{$|i\rangle$: otherwise}
\end{array}\right., 
\label{eq:holon}
\end{eqnarray}
$\tau^{(')}$ runs all the NN (diagonal-neighbor) sites of 
the site $i$, and $\mu^{(')}$ is a variational parameter 
controlling the binding of M and H between NN (diagonal-neighbor) 
sites. 
For TAL, we take account only of the NN M-H correlation:  
\begin{equation}
{\cal P}_Q(\mu)=(1-\mu)^{\hat{Q}}.  
\label{eq:pq-TAL}
\end{equation} 
In short, $\hat{Q}$ counts the number of M without NN H plus that 
of H without NN M.
The range of $\mu$ is limited to $0\le\mu\le 1$, whereas a subsidiary 
parameter $\mu'$ sometimes becomes negative.\cite{notemu'} 
The density of M (H) isolated from H (M) is reduced by $\mu$; 
for $\mu=(\mu'=)\ 0$, $\Psi_{\rm DH}^\mu$ is reduced to $\Psi_{\rm G}$, 
and M can move freely from H. 
In the other limit, $\mu=1$, M (H) cannot appear unless it is 
accompanied by at least one H (M) in the adjacent sites. 
Such a complete D-H bound state is not necessarily insulating, 
as we will see later. 
In Fig.~\ref{fig:dh-exp-pw}, we show the weight of ${\cal P}_Q(\mu,0)$ 
as a function of nearest M(D)-to-H distance $r$ for an intermediate 
value ($\mu=0.7$). 
\par 
 
\vspace{-0.2cm}
%
{\bf (2)} 
We introduce long-range D-H binding factors of a simple form, 
which is formally written as, 
\begin{equation}
{\cal P}_Q(\lambda)=
\prod_j\left\{\left[1-\left(1-f(r_j)\right)\tilde d_j\right]
              \left[1-\left(1-f(r_j)\right)h_j\right]\right\}, 
\label{eq:LRDH}
\end{equation}
where $\lambda$ is a variational parameter included in $f(r)$, 
which controls the size of an M-H pair. 
In ${\cal P}_Q(\lambda)$, we consider M-H correlation only 
between each M and its nearest H, and vice versa; $r$ ($\ge 1$) 
denotes the distance between such M and H in unit of lattice 
constant, and is measured by the stepwise or ``Manhattan" metric. 
This fashion suits the spirit of strong-coupling expansion, and 
is not identical with ordinary Jastrow factors, in which 
all the pairs are taken into account. 
For $f(r)$, on the analogy of a fermionic case,\cite{YS,Miyagawa} 
we assume two primitive decaying forms, imposing $f(1)=1$: 
\begin{equation}
f(r) = \left\{
\begin{array}{ll}
\displaystyle
\exp\left(-\frac{r-1}{\lambda_{\rm e}}\right), & \quad\mbox{(a) exponential} \\
\displaystyle
\frac{1}{r^{\lambda_{\rm p}}}, & \quad\mbox{(b) power}
\end{array}\right. 
\label{eq:f(r)}
\end{equation}
whose behavior is sketched in Fig.\ref{fig:dh-exp-pw}. 
We write $\Psi_{\rm DH}^{\rm exp}$ and $\Psi_{\rm DH}^{\rm pwr}$ 
for the wave functions using the factors of eqs.~(\ref{eq:f(r)}a) and 
(\ref{eq:f(r)}b), respectively. 
Compared with the exponential form (a), the power-law form (b) naturally 
has a tail for large $r$ for intermediate to large $\lambda_{\rm e}$ 
and $1/\lambda_{\rm p}$. 
When $\lambda_{\rm e}$ or $1/\lambda_{\rm p}\rightarrow\infty$, 
${\cal P}_Q(\lambda)$ becomes unity, namely, $\Psi_{\rm DH}$ is reduced 
to $\Psi_{\rm G}$. 
Meanwhile, in the limit of $\lambda_{\rm e}$ or 
$1/\lambda_{\rm p}\rightarrow 0$, 
${\cal P}_Q(\lambda)$ is reduced to ${\cal P}_Q(\mu=1,\mu'=0)$, 
indicating the complete D-H binding within the NN sites. 
Since the probability density of M-H pairs with distance 
$r$ is controlled mainly by $g$ for small $U/t$ and by $\lambda$ for 
intermediate and large $U/t$, the mean M-H distance relates to 
$\lambda_{\rm e}$ or $\lambda_{\rm p}$ for $U\gsim U_{\rm c}$. 
\par

{\bf (3)} 
Instead of such an a priori form of $f(r)$ as eq.~(\ref{eq:f(r)}), 
we optimize $f(r)$ in eq.~(\ref{eq:LRDH}) for every $r$ 
as variational parameters, using recently-developed optimization 
techniques.\cite{Umrigar}
In this case [${\cal P}_Q(f)$], the number of variational 
parameters becomes equivalent to the linear dimension of 
the system $L$: $g$ and $f(2)$-$f(L)$. 
We represent this optimized-$f(r)$ wave function 
as $\Psi_{\rm DH}^{\rm opt}$. 
As we will see later, parameters $f(r)$ with large $r$ ($\gsim 7$), 
are actually unnecessary for any value of $U/t$, because such 
long-distance D-H pairs rarely appear. 
Although ${\cal P}_Q(f)$ is naturally better than both 
${\cal P}_Q(\mu,\mu')$ and ${\cal P}_Q(\lambda)$, we can learn 
much from the comparison with various types of ${\cal P}_Q$.
\par

Having considered an attractive part of the long-range correlation 
factor, we should check the effect on a Mott transition 
of a repulsive part, which is known to be important for off half 
filling.\cite{YS}
To this end, we consider the form, 
$
\Psi_{\rm J}={\cal P}_{\rm J}\Psi_{\rm DH}, 
$
with a common repulsive Jastrow factor,  
\begin{eqnarray}
{\cal P}_{\rm J}(a,\kappa)
    &=& \prod_{j,r}\Bigl\{\left[1-\left(1-\eta(r)\right)
                                       \tilde d_j\tilde d_{j+r}\right]
\qquad\qquad\nonumber \\
    &&\qquad\times\left[1-\left(1-\eta(r)\right)h_jh_{j+r}\right]\Bigr\}, 
\label{eq:Jast}
\end{eqnarray}
in which we take account of the correlation between all the M-M 
and H-H pairs regardless of pair distance, as in ordinary Jastrow 
factors. 
Here, we adopt a simple power-law decaying form of $\eta(r)$ as,
\begin{equation}
\eta(r)=1-\frac{a}{r^\kappa} \qquad (r\ge 1), 
\label{eq:Jastrow}
\end{equation}
where $a$ and $\kappa$ are variational parameters with ranges 
$0\le a\le 1$ and $0\le\kappa\le\infty$. 
Correspondingly, for ${\cal P}_Q$ in $\Psi_{\rm DH}$, we employ the 
power-law decaying form $\Psi_{\rm DH}^{\rm pwr}$ in eq.~(\ref{eq:f(r)}b). 
Although the form of eq.~(\ref{eq:Jastrow}) may not be the best, 
it is sufficient to ascertain the importance of the repulsive Jastrow 
factor for the Mott transition. 
\par

Incidentally, we compare the above wave functions with that studied 
in related papers by Capello \etal:\cite{Capello} 
\begin{equation}
\Psi=\exp\left[-\frac{1}{2}
\sum_{i,j}v_{i,j}(n_i-1)(n_j-1)\right]{\cal P}_Q\Phi_0,
\label{eq:Capello}
\end{equation} 
in which they emphasized importance of the long-range Jastrow factor. 
Equation (\ref{eq:Capello}) already has an attractive D-H factor 
in the Jastrow (exponential) part, but at each ($i,j$) the weight 
of attractive D-H correlation is inseparably connected 
with that of repulsive D-D (H-H) correlation. 
For $U\gsim U_{\rm c}/2$, since the site occupation number is 
almost restricted to $0\le p\le 2$, the number operator is written as,
\begin{equation}
n_j=1+d_j-h_j, 
\label{eq:number}
\end{equation} 
using a doublon operator $d_i$ in the space of $p\le 2$: 
\begin{equation}
d_i|i\rangle = \left\{
\begin{array}{ll}
1\ |i\rangle, & \quad\mbox{$|i\rangle$: D} \\
0\ |i\rangle, & \quad\mbox{$|i\rangle$: otherwise}
\end{array}\right.. 
\label{eq:doublon}
\end{equation}
Thereby, the exponential part of eq.~(\ref{eq:Capello}) becomes, 
\begin{equation}
\exp\left[
-\frac{1}{2}\sum_{i,j}v_{i,j}\left(d_id_j+h_ih_j-d_ih_j-h_id_j\right) 
\right], 
\label{eq:content}
\end{equation} 
which shows the attractive factor is a reciprocal of the repulsive 
factor. 
Thus, to adjust the D-H binding effect independently of the D-D 
correlation, one is obliged to add a redundant ${\cal P}_Q$ term. 
In this context, $\Psi_{\rm J}$ have the advantage in distinguishing 
the effect of D-H binding factors from that of repulsive ones, 
although eq.~(\ref{eq:Capello}) and $\Psi_{\rm J}$ may work similarly. 
\par

We first optimize these trial functions, applying an optimization VMC 
scheme to systems with up to 1,600 particles ($L=40$). 
With the optimal parameters obtained, we calculate the expectation 
values of relevant physical quantities, using a conventional 
VMC method. 
With this procedure, we can obtain accurate variational results 
in most cases, except for statistical errors. 
In Appendix, we briefly explain some details of the VMC calculations 
carried out in this paper. 
\par

\section{Short-Range Doublon-Holon Factor\label{sec:results}}
In this section, we study the short-range D-H binding wave function 
QWF, which exhibits typical properties of D-H-binding types of wave 
functions. 
In \S\ref{sec:optim}, we study the energy of QWF to find out a Mott 
critical behavior, by comparing with extreme cases. 
In \S\ref{sec:number}, we discuss the site-occupation number versus 
$U/t$, in connection with experiments of a quantum gas microscope. 
In \S\ref{sec:Mott-QWF}, the existence of Mott transition is 
corroborated and its properties are studied by various quantities. 
\par

\subsection{Overall behavior of QWF's energy\label{sec:optim}} 
%
\begin{figure}[hob]
\vspace{-0.2cm}
\begin{center}
\includegraphics[width=7.5cm,clip]{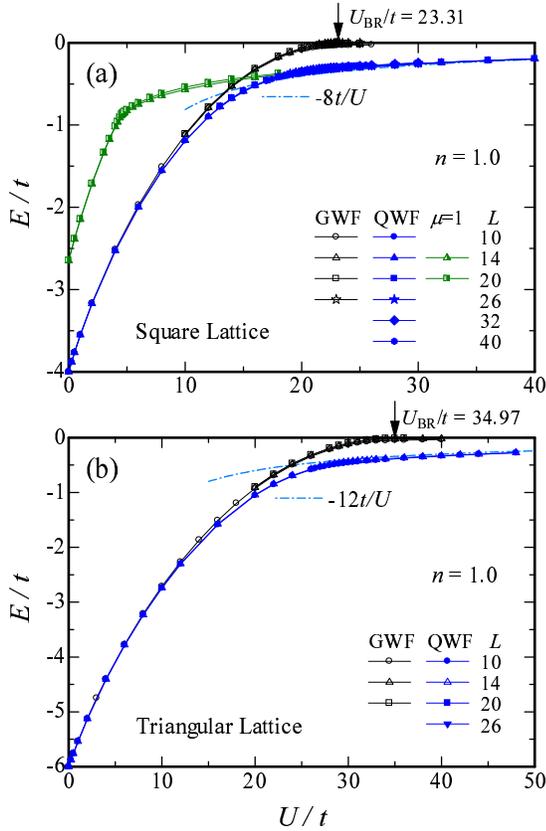}
\end{center}
\vskip -5mm
\caption{
(Color online) Energy expectation values of GWF (open symbols) 
and QWF (solid symbols) as a function of correlation strength 
$U/t$ for (a) the square and (b) the triangular lattices.
The result of strong-coupling expansion, eq.~(\ref{eq:effective}), 
is drawn by the dash-dotted lines. 
The critical values of Brinkman-Rice-type transitions in GWF 
are indicated by arrows on the upper axes. 
In (a), the data of QWF for ${\cal P}_Q(\mu=1, \mu'=0)$ are plotted 
with half-solid symbols (expressed as ``$\mu=1$"). 
The system-size dependence is inconspicuous in this scale. 
}
\label{fig:EvsU-mu}
\end{figure}
%
To begin with, we briefly check the optimized energy of QWF, 
eqs.~(\ref{eq:DH})-(\ref{eq:pq-TAL}). 
In Fig.~\ref{fig:EvsU-mu}(a), the total energies per site, $E/t$, 
are compared between GWF and QWF of ${\cal P}_Q(\mu,\mu')$ for SQL. 
For small $U/t$ (roughly $U<U_{\rm c}/2\sim 10t$), the improvement of 
$E/t$ on GWF is very small, indicating the D-H correlation plays 
a minor role for weakly interacting conductive states. 
For $U\gsim U_{\rm c}/2$, however, the curve of QWF gradually 
departs from that of GWF and approaches the extreme case of 
${\cal P}_Q(1,0)$, in which a doublon(s) and a holon(s) rigidly 
adhere to each other in the NN sites. 
Because such a state strongly suggests an insulator, a Mott transition 
is expected to occur near the value where the two curves join 
($U/t\sim 20$). 
The fact that the energy of QWF is considerably lowered from that 
of GWF in this regime indicates the D-H binding effect play a prominent 
part in the Mott physics. 
The transition arising in GWF is a continuous type, because $E/t$ 
vanishes as $\propto (1-U/U_{\rm BR})^2$ in the Bose fluid side. 
In contrast, QWF exhibits a first-order transition, as we will see 
in \S\ref{sec:Mott-QWF}.
Hence, the mechanisms of the two transitions are qualitatively 
distinct. 
\par 

The behavior in the regime of large $U/t$ should obey an effective 
Hamiltonian of eq.~(\ref{eq:model}) in the limit of $t/U\rightarrow 0$, 
which is led by a canonical transformation\cite{Harris} $e^{iS}$ with
\begin{equation}
S=\frac{it}{U}\sum_{\langle i,j\rangle}
\left(d_ib_i^\dag b_js_j+d_jb_j^\dag b_is_i
-s_ib_i^\dag b_jd_j-s_jb_j^\dag b_id_i
\right), 
\end{equation}
where $s_i$ is a projection operator of a singly-occupied site: 
\begin{equation}
s_i|i\rangle = \left\{
\begin{array}{ll}
1\ |i\rangle, & \quad\mbox{$|i\rangle$: singly occupied} \\
0\ |i\rangle, & \quad\mbox{$|i\rangle$: otherwise}
\end{array}\right.. 
\end{equation}
Then, we have an expression at $n=1$ in the space 
without D and H as, 
\begin{equation}
H_{\rm eff}=-\frac{4t^2}{U}\sum_{\langle i,j\rangle}s_is_j
=-2zN_{\rm s}\frac{t^2}{U}, 
\label{eq:effective}
\end{equation}
which is drawn as dash-dotted lines in Figs.~\ref{fig:EvsU-mu}(a) 
and 2(b). 
The energy of QWF well coincides with eq.~(\ref{eq:effective}) 
for a wide range of $U$ ($>U_{\rm c})$, meaning density fluctuation 
is properly introduced in the insulating regime. 
In Fig.~\ref{fig:EvsU-mu}(b), the same quantity is plotted 
for TAL.  
Because the behavior is qualitatively identical, henceforth, 
we address only SQL in most cases.
\par

\subsection{Site-occupation number and parity correlation
\label{sec:number}} 
%
\begin{figure}[hob]
\vspace{-0.2cm}
\begin{center}
\includegraphics[width=7.5cm,clip]{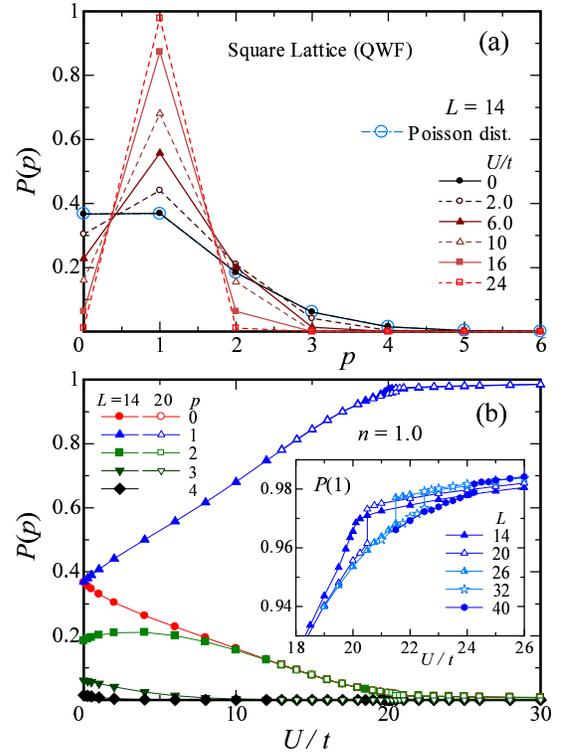}
\end{center}
\vskip -5mm
\caption{
(Color online) Ratio of the number of sites occupied by $p$ particles 
($N_p/N_{\rm s}$) calculated with QWF are shown,  
(a) as a function of $p$ for some $U/t$, and 
(b) as a function of $U/t$ for small $p$. 
In (a) the Poisson distribution eq.~(\ref{eq:Poisson}) is plotted 
with large open circles to confirm the VMC outcome for $U/t=0$. 
System-size dependence is negligible except for the Mott critical 
region, whose magnification for the case of $p=1$ is shown in the 
inset in (b) for several $L$. 
}
\label{fig:ndist-density}
\end{figure}
%
In connection with the Mott transition, a change in the distribution 
[$P(p)$] of site-occupation number $p$ have been directly observed 
in recent experiments of cold bosonic atom 
gases.\cite{number-Ger,number-Che} 
Since noninteracting bosons are randomly distributed to the lattice 
sites, the ratio of the number of sites occupied by $p$ particles ($N_p$) 
at $U/t=0$ should obey a Poisson distribution:
\begin{equation}
P(p)=\frac{N_p}{N_{\rm s}}=e^{-n}\frac{n^p}{p!},
\label{eq:Poisson}
\end{equation}
with $n=1$ at unit filling. 
As $U/t$ increases, however, $N_p$ with large $p$ rapidly decreases
to reduce $\langle H_U\rangle=U\langle D\rangle$, and for $U>U_{\rm c}$ 
a number squeezed state is considered to be realize, in which most sites 
are loaded with one particle, $P(1)\sim 1$. 
The $U/t$ dependence of $P(p)$ has been addressed using GWF\cite{BHM-GWF1}, 
a mean field approximation\cite{Lu} and QMC with analytic 
confirmations.\cite{Sansone} 
With these studies in mind, we discuss the results of $\Psi_{\rm DH}$. 
\par 

Figure \ref{fig:ndist-density}(a) shows the evolution of $P(p)$ for QWF, 
as $U/t$ is varied. 
The Poissonian is changed to a shape of a symmetric peak centered at $p=1$ 
at a relatively small value of $U/t$, and $P(p)$ with $p\ge 3$ almost 
vanishes; at $U/t=12$, $P(3)$ is less than $P(2)/100$ (see Table 
\ref{table:P(p)}). 
The $U/t$ dependence of $P(p)$ depends very slightly on the type 
of D-H correlation factor, as shown in Table \ref{table:P(p)}, 
where the values of $P(p)$ are compared among various correlation factors
for some values of $U/t$. 
Furthermore, the $U/t$ dependence shown in Fig.~\ref{fig:ndist-density}(b) 
is quantitatively consistent with a result of QMC [Fig.~1(b) in 
ref.~\citen{Sansone}]. 
Thus, in the Mott critical regime ($U\gsim U_{\rm c}/2$), we can safely 
consider the problem in the restricted space of $p\le 2$, similarly 
to Fermi systems. 
\par

\begin{table}
\caption{
Comparison of probability of site occupation $P(p)$ with $p=0$-$3$ 
among GWF and four types of $\Psi_{\rm DH}$ for three values of $U/t$ 
for $L=20$. 
For $U/t\ge 12$, $P(4)$ is less than $10^{-5}$ in every case. 
In the state column, SF and MI denote `superfluid' and `Mott insulating', 
respectively. 
Incidentally, $P(0)$ and $P(2)$ of GWF must completely vanish 
for $L=\infty$ in the MI regime. 
}
\label{t1}
\begin{center}
\begin{tabular}{c|c|c|c|c|c}
\hline
$\Psi$ & state & $P(0)$ & $P(1)$ & $P(2)$ & $P(3)$ \\
\hline\hline
\multicolumn{6}{l}{$U/t=0$\ \ (Poisson distribution)} \\
\hline
 -  & SF & 0.36788 & 0.36788 & 0.18394 & 0.06131 \\
\hline\hline
\multicolumn{6}{l}{$U/t=12$} \\
\hline
GWF  & SF & 0.14503 & 0.71147 & 0.14197 & 0.00153 \\
QWF  & SF & 0.12694 & 0.74717 & 0.12485 & 0.00104 \\
exp. & SF & 0.13053 & 0.73986 & 0.12869 & 0.00092 \\
pwr. & SF & 0.12742 & 0.74605 & 0.12565 & 0.00089 \\
opt. & SF & 0.12670 & 0.74755 & 0.12481 & 0.00095 \\
\hline\hline
\multicolumn{6}{l}{$U/t=19$} \\ 
\hline
GWF  & SF & 0.04958 & 0.90084 & 0.04956 & 0.00001 \\
QWF  & SF & 0.02982 & 0.94038 & 0.02979 & 0.00002 \\
exp. & MI & 0.01471 & 0.97057 & 0.01471 & 0.00000 \\
pwr. & SF & 0.02866 & 0.94269 & 0.02864 & 0.00001 \\
opt. & SF & 0.02855 & 0.94290 & 0.02854 & 0.00001 \\
\hline\hline
\multicolumn{6}{l}{$U/t=24$} \\ 
\hline
GWF  & MI & 0.00108 & 0.99783 & 0.00108 & 0.00000 \\
QWF  & MI & 0.01016 & 0.97968 & 0.01016 & 0.00000 \\
exp. & MI & 0.01014 & 0.97973 & 0.01013 & 0.00000 \\
pwr. & MI & 0.01035 & 0.97931 & 0.01035 & 0.00000 \\
opt. & MI & 0.01038 & 0.97925 & 0.01038 & 0.00000 \\
\hline
\end{tabular}
\end{center}
\label{table:P(p)}
\end{table}
As seen in Fig.~\ref{fig:ndist-density}(b), the ratio of singly-occupied 
sites (doublon and holon densities) increases (decrease) almost linearly 
with $U/t$ until near the critical point. 
The variation of $P(1)$ at the Mott critical point is as small as 1\% 
in QWF [see the inset of Fig.~\ref{fig:ndist-density}(b)], and 
local number fluctuation remains in some degree even in the Mott 
insulating phase. 
Thus, a number squeezed state is gradually approached as $U/t$ 
increases, and does not distinctly characterize a Mott insulating (MI) 
state. 
\par

%
\begin{figure}[hob]
\vspace{-0.2cm}
\begin{center}
\includegraphics[width=7.5cm,clip]{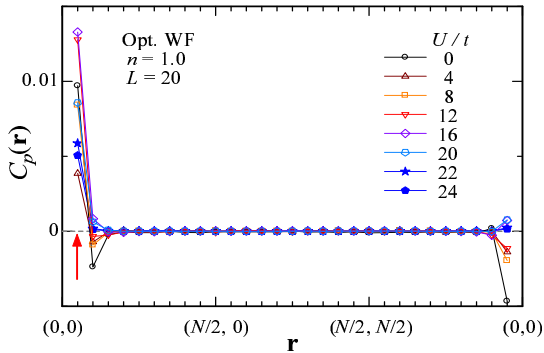}
\end{center}
\vskip -5mm
\caption{(Color online) 
Real-space parity correlation function for various values of $U/t$ 
calculated with long-range D-H binding wave function 
$\Psi_{\rm DH}^{\rm opt}$ along path shown on abscissa. 
The arrow indicates the nearest-neighbor site. 
The data in the Mott insulating regime are indicated by solid symbols. 
}
\label{fig:C-p}
\end{figure}
%
Recent development of single atom-single lattice site imaging technique 
(a quantum gas microscope) enables one to directly observe the parity 
(even or odd) of the occupied particle number $p$ at each 
site;\cite{Bakr,Sherson} the parity of the site $j$ is written as, 
\begin{equation}
\bar p_j=\frac{1}{2}\left[1-(-1)^{n_j}\right].
\label{eq:parity}
\end{equation}
Thus, the correlation function of the parity, 
\begin{equation}
C_{\bar p}(j,\ell)=\langle \bar p_j\bar p_\ell\rangle
           -\langle \bar p_j\rangle\langle \bar p_\ell\rangle
\end{equation}
becomes a quantity directly measured by experiments.\cite{Kapit} 
In Fig.~\ref{fig:C-p}, we show $C_{\bar p}(j,\ell)$ for various 
values of $U/t$ using the long-range $\Psi_{\rm DH}$ with ${\cal P}_Q(f)$. 
Because $C_{\bar p}(j,\ell)$ is closely related to the usual 
density correlation function,
\begin{equation}
N(j,\ell)=\langle n_jn_\ell\rangle
           -\langle n_j\rangle\langle n_\ell\rangle, 
\label{eq:dcf}
\end{equation}
the magnitude of $C_{\bar p}(j,\ell)$ rapidly decays, as $|{\bf r}|$ 
increases.
For large $U/t$, in particular in the insulating regime, the parity 
correlation is almost restricted to the nearest-neighbor sites, 
suggesting insignificance of the long-range part of correlation. 
In this regime, the parity operator eq.~(\ref{eq:parity}) is written 
by the projection operators eqs.~(\ref{eq:holon}) and (\ref{eq:doublon}) 
as,
\begin{equation}
\bar p_j=1-d_j-h_j, 
\end{equation}
and $n_j$ by eq.~(\ref{eq:number}), so that $C_{\bar p}(j,\ell)$ is 
written with $N(j,\ell)$ and the doublon-holon correlation function: 
\begin{equation}
C_{\rm DH}(j,\ell)=\langle d_jh_\ell\rangle
           -\langle d_j\rangle\langle h_\ell\rangle,
\end{equation}
as, 
\begin{equation}
C_{\bar p}(j,\ell)=N(j,\ell)+4C_{\rm DH}(j,\ell). 
\end{equation}
Thus, the density and D-H correlation functions are fundamental 
also in analyzing quantum gas microscope experiments 
near the Mott transition. 
\par

\subsection{Mott transition in QWF\label{sec:Mott-QWF}} 
\begin{figure}[hob]
\vspace{-0.2cm}
\begin{center}
\includegraphics[width=8.0cm,clip]{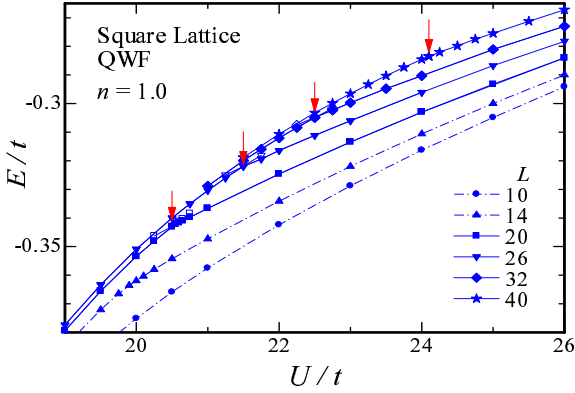}
\end{center}
\vskip -5mm
\caption{
(Color online) Magnification of total energy of QWF 
[Fig.~\ref{fig:EvsU-mu}(a)] near Mott critical 
points as function of interaction strength for several values of $L$. 
The arrows indicate the transition points ($U_{\rm c}/t$).
Open symbols near $U_{\rm c}/t$ denote the data of metastable states. 
}
\label{fig:EvsUm-c}
\end{figure}

\begin{figure}[hob]
\vspace{-0.2cm}
\begin{center}
\includegraphics[width=7.5cm,clip]{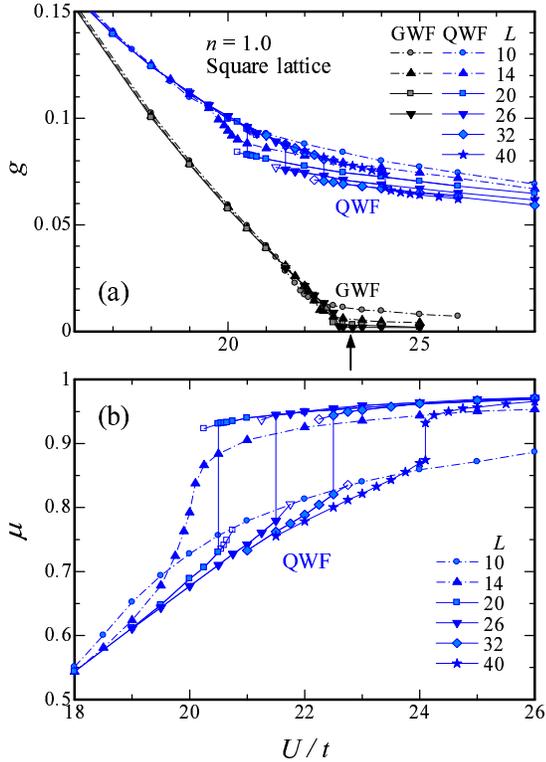} 
\end{center}
\vskip -5mm
\caption{(Color online) 
Optimized main variational parameters in QWF for square lattice, 
(a) Gutzwiller parameter and 
(b) nearest-neighbor D-H parameter, 
near Mott critical values for some $L$. 
In (a), corresponding values of GWF are added with an arrow indicating 
$U_{\rm BR}$. 
Open symbols near $U_{\rm c}/t$ denote the values of metastable states. 
}
\label{fig:para-c}
\end{figure}

\begin{figure}[hob]
\vspace{-0.2cm}
\begin{center}
\includegraphics[width=7.0cm,clip]{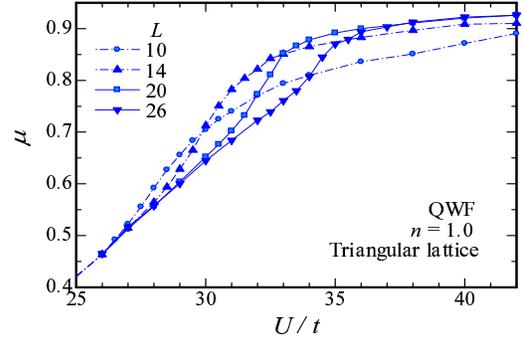} 
\end{center}
\vskip -5mm
\caption{(Color online) 
Optimized nearest-neighbor D-H parameter near Mott critical 
values in QWF for triangular lattice. 
Four system sizes ($L$) are compared. 
}
\label{fig:paramum}
\end{figure}

Now, we analyze the superfluid-insulator transition in QWF more 
in detail. 
When we carefully watch the behavior of $E/t$ in a magnified 
figure (Fig.~\ref{fig:EvsUm-c}), we notice a cusp for each $L$ 
with $L\ge 20$. 
In each side of the cusp, we can find another local minimum 
(a metastable point) of $E/t$, which is smoothly extrapolated 
from the curve in the other side of the cusp, suggesting 
a first-order transition. 
This is supported by the existence of a discontinuity at the 
cusp point in the optimized variational parameters, as shown 
in Fig.~\ref{fig:para-c}. 
Critical values thus obtained are listed in Table \ref{table:Uc};
the system-size dependence of $U_{\rm c}/t$ will be discussed 
in \S\ref{sec:LRMott}. 
A clear sign of a first-order transition is not observed for 
small systems ($L\le 14$ in this case), similarly to fermionic 
systems.\cite{YOT}
For TAL, a clear discontinuity does not appear even for $L=26$, 
as shown in Fig.~\ref{fig:paramum} for $\mu$, and $E/t$ has 
very broad minimum at $U\sim U_{\rm c}$; the tendency toward 
a continuous transition is stronger for TAL, which tendency 
is similar to those of frustrated metallic states in fermionic 
models.\cite{YOT}
\par

%
In contrast to GWF, which has $g=0$ for $U>U_{\rm BR}$, QWF has 
finite $g$ even for $U>U_{\rm c}$, indicating the existence 
of density fluctuation even in the MI regime. 
The nearest-neighbor D-H parameter $\mu$ exhibits a large discontinuity 
at $U_{\rm c}$, and becomes close to 1 for $U>U_{\rm c}$; 
the D-H binding becomes firm in the insulating regime. 
Note that, as we will see, the Mott transition is induced 
by the collaboration of the suppression of onsite density fluctuation 
by $g$ and the D-H binding effect by ${\cal P}_Q$. 
\par

\begin{figure}[hob]
\vspace{-0.2cm}
\begin{center}
\includegraphics[width=8.5cm,clip]{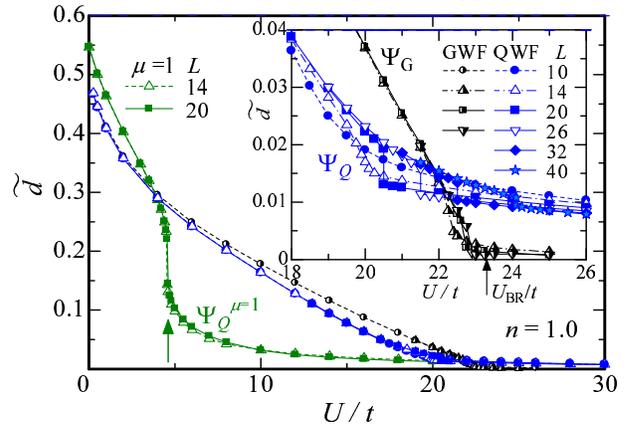} 
\end{center}
\vskip -5mm
\caption{(Color online) 
Comparison of expectation values of onsite correlation operator $D$
[eq.~(\ref{eq:D})] per siteon the square lattice among GWF, QWF and 
completely D-H bound state $\Psi_Q^{\mu=1}$ (discussed later). 
This quantity is substantially doublon density $d$ for $U/t\gsim 10$, 
and is half the onsite density fluctuation $\sigma_o^2$. 
The arrow on the curve of $\Psi_Q^{\mu=1}$ indicates the Mott 
critical point.
The inset shows the magnification near the Mott critical points 
for GWF and QWF. 
The symbols in the inset are common to those in the main panel. 
} 
\label{fig:DvsUm}
\end{figure}
%
To see the properties of this first-order transition, we discuss 
some relevant quantities. 
We start with the expectation value of onsite correlation operator $D$, 
eq.~(\ref{eq:D}). 
Especially at unit filling ($\langle n_j\rangle=\langle n_j\rangle^2$=1), 
this quantity per site, $\tilde d=\langle D\rangle/N_{\rm s}$, 
coincides with half of the onsite density fluctuation or variance, 
\begin{eqnarray}
\sigma^2_{\rm o}&\equiv& N(j,j)
=\langle n_j^2\rangle-\langle n_j\rangle^2 
\label{eq:N00}
\\
&=&\langle n_j^2\rangle-\langle n_j\rangle
=2\tilde d \qquad(n=1).
\label{eq:N00uf}
\end{eqnarray}
Substituting eq.~(\ref{eq:number}) in eq.~(\ref{eq:N00uf}), we have 
a relation, $\tilde d=d$, valid for $U\gsim U_{\rm c}/2$, where $d$ 
is the doublon density. 
Thus, $d$ is a key indicator of Mott transitions like for fermionic 
systems, so that we regard $d$ as an order parameter of Mott transitions 
also for bosons. 
In Fig.~\ref{fig:DvsUm}, we show the $U/t$ dependence of $\tilde d$ 
for GWF and QWF, and its magnification near $U_{\rm c}$ in the inset. 
For each value of $L$, $\tilde d$ of QWF decreases with $U/t$ and 
exhibits a discontinuity at each $U_{\rm c}/t$ ($L\ge20$). 
Conversely, the kinetic energy $E_t=\langle H_t\rangle$ increases 
with a discontinuity at $U_{\rm c}$ (not shown). 
This energetics meets the ordinary criterion of Mott transitions: 
The energy is stabilized by lowering $E_U$ at the cost of $E_t$ 
for $U>U_{\rm c}$. 
The doublon density $\tilde d$ for QWF remains finite in the insulating 
regime, because local density fluctuation is permitted 
and transient D-H pairs exist within neighboring sites [see also $P(2)$ 
in Table \ref{table:P(p)}]. 
This behavior contrasts with that of GWF, for which $\tilde d$ vanishes 
for $U>U_{\rm BR}$. 
Note that this quantity will be measured as a function of $U/t$ 
by experiments like quantum gas microscopes. 
\par

\begin{figure}[hob]
\vspace{-0.2cm}
\begin{center}
\includegraphics[width=7.0cm,clip]{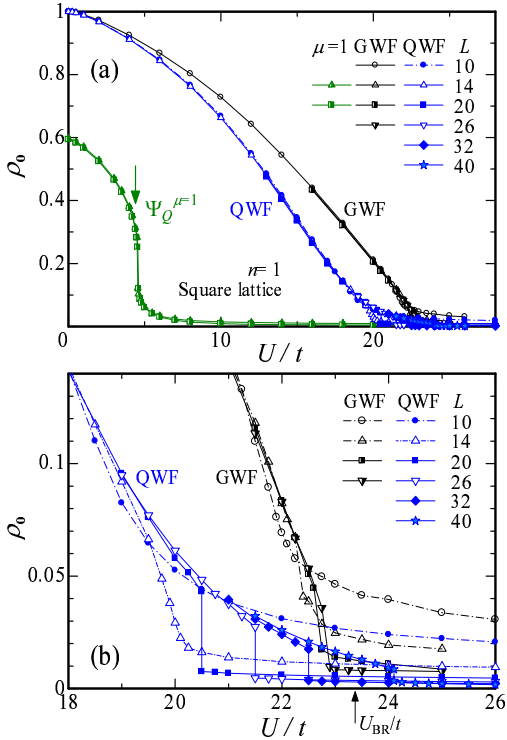} 
\end{center}
\vskip -5mm
\caption{(Color online) 
(a) Condensate fraction as a function of $U/t$ for some values 
of $L$, calculated with three functions, GWF, QWF, and 
completely D-H bound function, $\Psi_Q^{\mu=1}$ (discussed later). 
The arrow on $\Psi_Q^{\mu=1}$ indicates the Mott critical point. 
(b) Magnification of the Mott critical region for GWF and QWF.   
}
\label{fig:cor0-cc} 
\end{figure}
%
Next, we consider $U/t$ dependence of the condensate fraction, 
\begin{equation}
\rho_{\bf 0}=n({\bf 0})/N_{\rm s},
\label{eq:rho_0}
\end{equation} 
with $n({\bf 0})$ being the ${\bf k}={\bf 0}$ element of the momentum 
distribution function: 
\begin{equation}
n({\bf k})=\langle b_{\bf k}^\dag b_{\bf k}\rangle
=\frac{1}{N_{\rm s}}\sum_{j,\ell}e^{i{\bf k}\cdot{\bf r_\ell}}
        \langle b_{j+\ell}^\dag b_j\rangle. 
\label{eq:nk}
\end{equation} 
In Fig.~\ref{fig:cor0-cc}, we plot $\rho_{\bf 0}$ calculated with 
GWF and QWF. 
As $U/t$ increases, the condensate fraction diminishes from the value 
of free bosons $\rho_{\bf 0}=1$, and vanishes at $U=U_{\rm c}$ 
as the order of $n/N_{\rm s}$ in the non-superfluid states 
for all the wave functions. 
Thus, superfluidity vanishes at $U=U_{\rm c}$. 
As shown in Fig.~\ref{fig:cor0-cc}(b), the discontinuities appear 
for QWF with $L\ge 20$ in accordance with other quantities. 
Recently, the condensate fraction has been actually observed 
by experiments of cold atoms.\cite{Mun,Spielman2} 
\par

\begin{figure}[hob]
\vspace{-0.2cm}
\begin{center}
\includegraphics[width=7.0cm,clip]{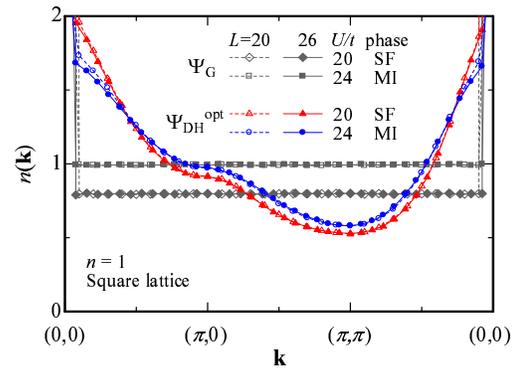} 
\end{center}
\vskip -5mm
\caption{(Color online) 
The momentum distribution function, eq.~(\ref{eq:nk}), are compared 
between GWF ($\Psi_{\rm G}$) and optimizing wave function 
($\Psi_{\rm DH}^{\rm opt}$), and between superfluid (SF) and Mott 
insulating (MI) phases. 
The value of the coherent wave number {\bf k}={\bf 0} is omitted. 
Two system sizes are used. 
The behavior of $n({\bf k})$ in the other D-H binding wave functions 
is similar to that of $\Psi_{\rm DH}^{\rm opt}$. 
}
\label{fig:nk}
\end{figure}
%
Here, we discuss ${\bf k}\ne{\bf 0}$ elements of $n({\bf k})$, because 
it is a directly-observed quantity by cold-atom experiments. 
Figure \ref{fig:nk} shows $n({\bf k})$ obtained by the VMC calculations 
under some conditions. 
Because GWF gives the result identical with that of the Gutzwiller 
approximation in fermionic cases,\cite{GA} $n({\bf k})$ is constant 
for ${\bf k}\ne{\bf 0}$ in both superfluid and insulating phases. 
On the other hand, wave functions with D-H factors yields dispersive 
$n({\bf k})$, owing to the effect of density fluctuation. 
In Fig.~\ref{fig:nk}, we draw $n({\bf k})$ of the best D-H binding 
wave function in this study $\Psi_{\rm DH}^{\rm opt}$. 
It is marked that the difference is small between the superfluid 
($U/t=20$) and insulating states ($U/t=24$). 
As ${\bf k}$ leaves the $\Gamma (={\bf 0})$ point, $n({\bf k})$ 
decreases in any direction, but its decrement is larger 
in the $\Gamma\rightarrow(\pi,\pi)$ direction than 
in the $\Gamma\rightarrow(\pi,0)$ direction. 
This direction-dependent dispersion near the Mott critical point is 
actually observed as a cross-like intensity in absorption images 
of cold atoms, for example in Figs.~2f and 2g of ref.~\citen{Greiner}. 
This topic was previously argued by the perturbative correction to 
the Gutzwiller solution.\cite{Schroll}
\par

%
\begin{figure*}[!t]
\vspace{0.2cm}
\begin{center}
\includegraphics[width=15.0cm,clip]{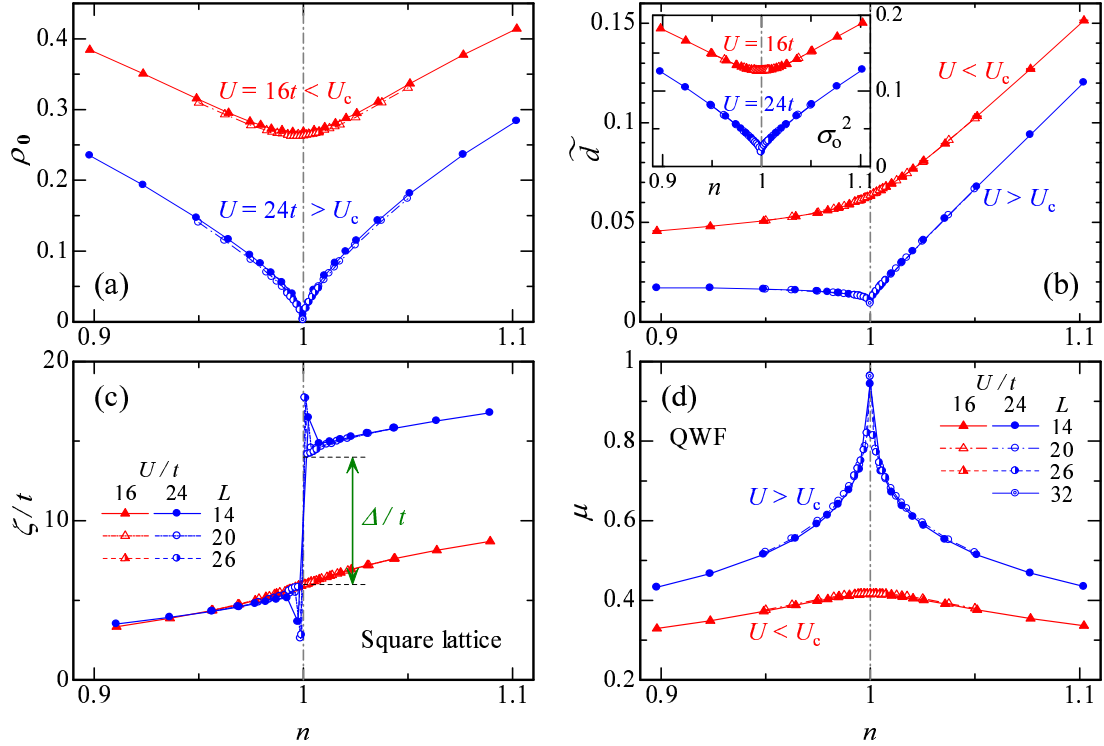} 
\end{center}
\vskip -5mm
\caption{(Color online) 
Particle-density dependence of various quantities are compared 
between at $U/t=16$ [a superfluid (SF) state for $n=1$], and 
at $U/t=24$ [a Mott insulating (MI) state for $n=1$], calculated with 
QWF on the square lattices of some system sizes $L\times L$. 
(a) Condensate fraction given by $n({\bf 0})/N_{\rm s}$. 
(b) The main panel shows the substantial doublon density, 
$\langle D\rangle/N_{\rm s}$, and 
the inset the onsite density fluctuation $\sigma_{\rm o}^2=N(j,j)$ 
defined by eq.~(\ref{eq:N00}). 
For $n\ne 1$, the relation $\sigma_{\rm o}^2=2\tilde d$ 
[eqs.~(\ref{eq:N00}) and (\ref{eq:N00uf})] does not hold. 
(c) Chemical potential, eq.~(\ref{eq:cp}), actually estimated 
from the finite difference of $E/t$ with respect to $n$. 
The arrow indicates the Mott gap for $U/t=24$ at $n=1$.
(d) Optimized nearest-neighbor D-H binding parameter. 
}
\label{fig:Qvsn-cc}
\end{figure*}
%
Finally, to corroborate a superfluid-insulator transition  
at unit filling, we compare particle-density ($n$) dependence 
of various quantities in the vicinity of $n=1$\cite{noteDH} 
between $U<U_{\rm c}$ and $U>U_{\rm c}$. 
Let us start with the condensate fraction $\rho_0$, which is plotted 
in Fig.~\ref{fig:Qvsn-cc}(a). 
For $U=16t$ $(<U_{\rm c})$, $\rho_{\bf 0}$ becomes minimum at $n=1$, 
but preserves a finite magnitude, whereas for $U/t=24$ $(>U_{\rm c}/t)$, 
$\rho_{\bf 0}$ decreases as $n$ approaches 1, and almost vanishes at 
$n=1$. 
The sign of second derivative $\partial^2\rho_{\bf 0}/\partial n^2$ 
becomes different between the two cases near unit filling. 
Thus, a MI state appears only at $n=1$ and $U>U_{\rm c}$. 
This is supported by the similar behavior of the onsite density 
fluctuation $\sigma_{\rm o}^2$, eq.~(\ref{eq:N00}), as shown 
in the inset of Fig.~\ref{fig:Qvsn-cc}(b).
The order parameter of Mott transitions, $\tilde d$, depicted in 
the main panel of Fig.~\ref{fig:Qvsn-cc}(b) monotonically 
increases for $U<U_{\rm c}$, whereas it sharply decreases at $n=1$ 
for $U>U_{\rm c}$, suggesting a Mott transition. 
Figure \ref{fig:Qvsn-cc}(c) represents the chemical potential,
\begin{equation}
\zeta\left(N-\frac{1}{2}\right)=E(N)-E(N-1)
                      =\frac{\partial E}{\partial n}
\label{eq:cp}
\end{equation}
as a function of $n$. 
Although for $U<U_{\rm c}$, $\zeta/t$ is always a smooth function 
of $n$, for $U>U_{\rm c}$, it has a discontinuity corresponding 
to the Mott gap, broadly estimated as $\Delta/t=8.2$ for $U/t=24$ 
on SQL, and 17.6 for $U/t=40$ on TAL (figure not shown).
This is a direct evidence of the Mott transition. 
\par

To recognize the importance of the D-H binding effect on the 
Mott transition, we compare, in Fig.~\ref{fig:Qvsn-cc}(d), 
the optimized D-H binding parameter $\mu$ in QWF between the two 
phases. 
In the two phases, $\mu$ is symmetric with respect to $n=1$, and has 
a maximum at $n=1$, but the magnitude is distinct.
For $U=16t$, $\mu$ slowly varies with $n$ and is still small 
at $n=1$, whereas for $U=24t$, $\mu$ anomalously increases as $n$ 
approaches 1, and almost reaches 1 as $L$ increases. 
Thus, the D-H binding effect is significantly enhanced in the very 
vicinity of the insulating state.
\par

\section{Long-Range Correlation Factors\label{sec:long-range}}
In this section, we discuss the effect of long-range correlation 
factors, which we have disregarded in the preceding section. 
In \S\ref{sec:LRD-H} and \S\ref{sec:LRMott}, we focus on 
the properties of D-H attractive factors. 
In \S\ref{sec:LRD-D} we study the effect of additional D-D and H-H 
repulsive factors.
\par 

\subsection{Optimized D-H attractive factors\label{sec:LRD-H}}
\begin{figure}[hob]
\vspace{-0.2cm}
\begin{center}
\includegraphics[width=8.5cm,clip]{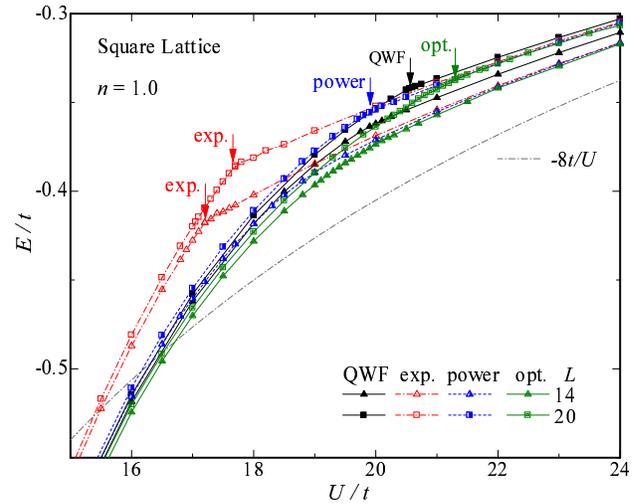} 
\end{center} 
\vskip -5mm
\caption{(Color online) 
The total energies are compared near the Mott critical points
among the wave functions with four kinds of D-H attractive factors: 
a short-ranged type (QWF) and three long-range types (exponentially- 
and power-law-decaying and completely-optimizing types). 
For clarity, only the data of two system sizes are plotted. 
The Mott critical point for each function ($L=20$, and $14$ 
for $\Psi_{\rm DH}^{\rm exp}$) are indicated by arrows. 
The value of strong-coupling expansion ($-8t/U$) is added. 
}
\label{fig:EvsUm-cc}
\end{figure} 
%
To begin with, we compare the minimized energy among QWF and 
the wave functions with three long-range D-H attractive factors, 
exponentially decaying ($\Psi_{\rm DH}^{\rm exp}$), power-law 
decaying ($\Psi_{\rm DH}^{\rm pwr}$), and completely optimizing 
($\Psi_{\rm DH}^{\rm opt}$) types, introduced in \S\ref{sec:wf}. 
As will be discussed in \S\ref{sec:LRMott}, a Mott transition occurs 
in each wave function near $U_{\rm c}$ of QWF ($\sim 20t$). 
For sufficiently large and small $U$, compared with $U_{\rm c}$, 
the total energies of the four functions are close to one another. 
As shown in Fig.~\ref{fig:EvsUm-cc}, some difference appears 
in the Mott critical region. 
Of course, $E^{\rm opt}$ ($E$ for $\Psi_{\rm DH}^{\rm opt}$) is 
always the lowest. 
For $U/t\gsim 21$, the energies of three long-range D-H wave 
functions are broadly analogous and somewhat improved over 
$E^{\rm QWF}$ for each system size. 
On the other hand, for $U/t\lsim 20$, $E^{\rm exp}$ becomes 
clearly higher than $E/t$ of the other D-H wave functions; 
$E^{\rm pwr}$ is also slightly higher than $E^{\rm QWF}$. 
Now, we are aware that the frequently-used QWF yields a relatively 
good result, especially for $U<U_{\rm c}$, despite its simplicity. 
We analyze these features of $E/t$ by comparing the optimized 
forms of correlation factors $f(r)$ among the D-H wave functions, 
in the following. 
\par

\begin{figure}[hob]
\vspace{-0.2cm}
\begin{center}
\includegraphics[width=8.5cm,clip]{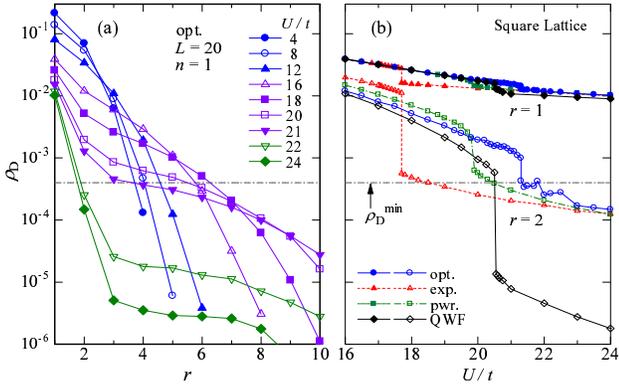} 
\end{center} 
\vskip -5mm
\caption{(Color online) 
The appearance probability of a doublon the distance from which 
to the nearest holon is $r$ is plotted, 
(a) for the completely optimizing wave function $\Psi_{\rm DH}^{\rm opt}$
($U_{\rm c}/t\sim 21.3$), versus $r$ and for some values of $U/t$, and 
(b) for the four D-H wave functions versus $U/t$ for $r=1$ and 2. 
The threshold value $\rho_{\rm D}^{\rm min}=0.0004$ is indicated 
by a dash-dotted line in both panels. 
}
\label{fig:rho-D-cc} 
\end{figure}
%
First, we discuss the effective range of $f(r)$. 
We represent the distance from a doublon (multiplon) to its nearest 
holon simply by $r$ here, and the probability that a site is occupied 
by a doublon of $r$ by $\rho_{\rm D}(r)$, which satisfies the relation, 
\begin{equation}
\sum_{r=1}^L\rho_{\rm D}(r)=\tilde d. 
\end{equation}
In Fig.~\ref{fig:rho-D-cc}(a), we plot $\rho_{\rm D}(r)$ of 
$\Psi_{\rm DH}^{\rm opt}$ in a logarithmic scale. 
Because $\rho_{\rm D}(r)$ is a monotonically decreasing function of $r$, 
it is convenient to define a threshold value $\rho_{\rm D}^{\rm min}$, 
which broadly gives the effective range of $f(r)$ by, 
\begin{equation}
\rho_{\rm D}(r)>\rho_{\rm D}^{\rm min}. 
\label{eq:range}
\end{equation}
The contribution from $\rho_{\rm D}(r)<\rho_{\rm D}^{\rm min}$ should 
be negligible, namely, the corresponding particle configurations should 
appear very rarely. 
Thus, the optimized values of $f(r)$ for such distant $r$ are 
unreliable and insignificant. 
Here, we set $\rho_{\rm D}^{\rm min}=100/250,000$ on the basis of 
accuracy in the VMC calculations. 
In a weakly interacting regime ($U/t\lsim 12$), the weight of 
$\rho_{\rm D}(r)$ concentrates in $r\lsim 4$, and virtually vanishes 
($\rho_{\rm D}(r)<\rho_{\rm D}^{\rm min}$) for larger $r$, because 
multiplons and holons are crowded. 
In the superfluid regime, $\rho_{\rm D}(r)$ comes to decrease slowly 
as $U/t$ increases, and extend the valid range to $r\lsim 6$ 
near $U_{\rm c}/t$. 
Meanwhile, in the insulating regime, $\rho_{\rm D}(r)$ immediately 
decays to have substantial weight only on $r=1$ or at most 2. 
This contrastive feature of the effective range clearly implies 
that the nature of D-H binding effect changes at $U_{\rm c}$. 
This effective range of $f(r)$ is closely related to the D-H binding 
length, $\xi_{\rm dh}$, introduced in \S\ref{sec:picture}. 
Making a similar analysis with the condition (\ref{eq:range}) 
for the other wave functions, we determine the effective range 
of $f(r)$ for various values of $U/t$ for each function. 
\par

\begin{figure}[hob]
\vspace{-0.2cm}
\begin{center}
\includegraphics[width=8.0cm,clip]{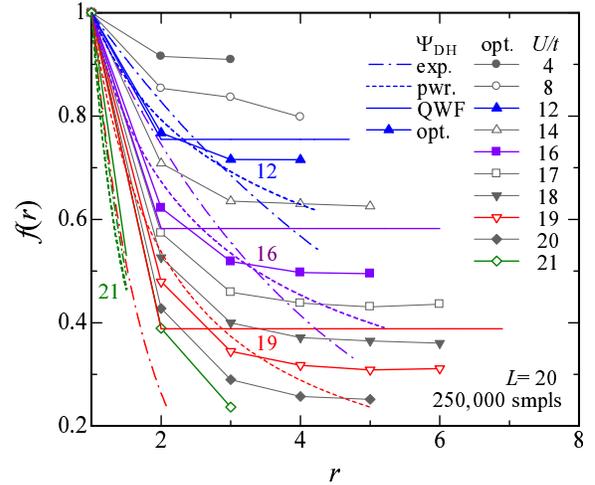} 
\end{center} 
\vskip -5mm
\caption{(Color online) 
The optimized D-H attractive factors are compared among four wave 
functions versus nearest D-to-H distance $r$. 
Except for $\Psi_{\rm DH}^{\rm opt}$, the data for $U/t=12$,~16,~19 
and 21 are shown for clarity. 
The range of $r$ is restricted according to the condition (\ref{eq:range}). 
The data of $U/t=19$ for $\Psi_{\rm DH}^{\rm exp}$ and of $U/t=21$ 
except for $\Psi_{\rm DH}^{\rm opt}$ are in the insulating phase. 
For QWF, we disregard the contribution of $\mu'$. 
}
\label{fig:paravsr-cc} 
\end{figure}
%
\begin{table}
\begin{center}
\caption{
Comparison of appearance probability and weight of D-H factors 
for $r=1$-3 among four wave functions in insulating regime ($U/t=22$) 
for $L=20$. 
}
\label{table:D-H}
\begin{tabular}{l|l|l|l|l|l|l}
\hline
\multicolumn{1}{c|}{\mbox{$\Psi_{\rm DH}$}} &
\multicolumn{1}{c|}{\mbox{$\rho_{\rm D}(1)$}} &
\multicolumn{1}{c|}{\mbox{$\rho_{\rm D}(2)$}} & 
\multicolumn{1}{c|}{\mbox{$f(2)$}} &
\multicolumn{1}{c|}{\mbox{$\rho_{\rm D}(3)$}} & 
\multicolumn{1}{c|}{\mbox{$f(3)$}} &
\multicolumn{1}{c}{\mbox{$E/t$}} \\ 
       & $\times 10^{-2}$ & $\times 10^{-4}$ & & $\times 10^{-5}$ & & \\
\hline
optim. & 1.18 & 2.54  & 0.263 & 2.57 & 0.090 & -0.3282 
\\                            
power  & 1.17 & 2.08  & 0.247 & 3.80 & 0.109 & -0.3279
\\                            
exp.   & 1.15 & 1.75  & 0.236 & 0.84 & 0.056 & -0.3272
\\                            
QWF    & 1.05 & 0.052 & 0.050 & 0.64 & 0.050 & -0.3246
\\
\hline
\end{tabular}
\vskip -5mm
\end{center}
\end{table}
%
Figure \ref{fig:paravsr-cc} compares the behavior of optimized $f(r)$ 
among the four wave functions within the effective range thus determined. 
In the conductive regime, $f(r)$ in $\Psi_{\rm DH}^{\rm opt}$ rapidly 
decreases for $r\lsim 3$, but becomes almost constant for $r\gsim 3$. 
Namely, the D-H binding is effective only for $r\lsim 3$, and a doublon 
is released from the bondage of holons for $r\gsim 3$. 
Because $f(r)$ of QWF is constant for $r\ge 3$, the behavior 
for large $r$ is analogous to $f(r)$ in $\Psi_{\rm DH}^{\rm opt}$. 
In contrast, $f(r)$ in $\Psi_{\rm DH}^{\rm pwr}$ and especially in 
$\Psi_{\rm DH}^{\rm exp}$ continues decreasing to zero as $r$ increases, 
namely, the effective range of D-H binding is too long, compared with 
$f(r)$ in $\Psi_{\rm DH}^{\rm opt}$. 
This is a cause of the unexpected good (unsatisfactory) result of QWF 
($\Psi_{\rm DH}^{\rm exp}$). 
In the MI regime, $f(r)$ in $\Psi_{\rm DH}^{\rm opt}$ rapidly decays 
with $r$, and the effective range is limited to at most $r=2$, 
as mentioned [Fig.~\ref{fig:rho-D-cc}(a)]. 
Regarding energy improvement, the appearance probability of 
nearest-neighbor D-H pairs seems primarily important; as listed 
in Table \ref{table:D-H}, $\rho_{\rm D}(1)$ is more than 10\% smaller 
in QWF than in the long-range wave functions. 
In addition, the role of $f(2)$ is not negligible. 
In Fig.~\ref{fig:rho-D-cc}(b), $\rho_{\rm D}(r)$ for $r=1$ and 2 
are compared in a logarithmic scale. 
For $U>U_{\rm c}$, the three long-range wave functions have 
similar values of $\rho_{\rm D}(2)$, whereas the values of QWF 
are roughly two orders of magnitude smaller. 
This is directly reflected in $f(2)$, as shown in Table \ref{table:D-H}. 
No such great differences in $\rho_{\rm D}(r)$ and $f(r)$ can be seen 
for $r\ge 3$ among the four wave functions [see also 
Fig.~\ref{fig:paraf3vsUm-cc}(b)]. 
\par

\subsection{Mott transitions in D-H attractive factors\label{sec:LRMott}}
\begin{figure}[hob]
\vspace{-0.2cm}
\begin{center}
\includegraphics[width=8.0cm,clip]{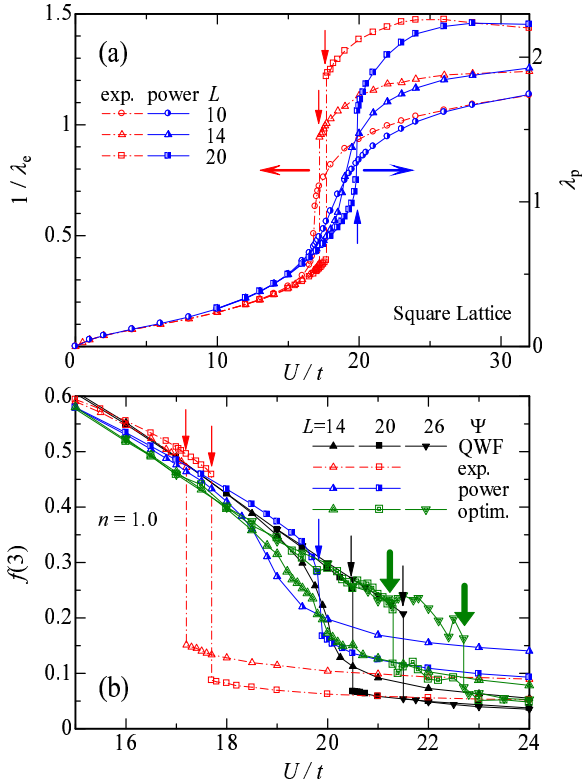} 
\end{center} 
\vskip -5mm
\caption{(Color online) 
(a) The optimized parameters controlling D-H correlation length, 
$1/\lambda_{\rm e}$ in $\Psi_{\rm DH}^{\rm exp}$ and $\lambda_{\rm p}$ 
in $\Psi_{\rm DH}^{\rm pwr}$ are plotted for $L=10$-20. 
(b) The weight of D-H attractive factors $f(r)$ for $r=3$ is compared 
near the Mott critical points among the four D-H binding wave functions. 
In both panels, the critical points are indicated by vertical arrows, 
in case they are manifest. 
The thick arrows for $\Psi_{\rm DH}^{\rm opt}$ in (b) means large 
ambiguity in $U_{\rm c}/t$. 
}
\label{fig:paraf3vsUm-cc}
\end{figure}
%
In this subsection, we consider other properties of long-range 
wave functions as to the Mott transition. 
\par 

We start with the parameters which controls the range of D-H 
attractive correlation, namely $\lambda_{\rm e}$ 
in $\Psi_{\rm DH}^{\rm exp}$ and $\lambda_{\rm p}$ in 
$\Psi_{\rm DH}^{\rm pwr}$ (Fig.~\ref{fig:dh-exp-pw}). 
Their optimized values are shown in Fig.~\ref{fig:paraf3vsUm-cc}(a). 
Both $\lambda_{\rm e}$ and $1/\lambda_{\rm p}$ abruptly increases 
at $U/t=17$-20, namely, the D-H correlation range becomes short. 
In particular, clear jumps exist for $L\ge 14$ 
in $1/\lambda_{\rm e}$ and $L=20$ in $\lambda_{\rm p}$, 
as indicated by arrows. 
In Fig.~\ref{fig:paraf3vsUm-cc}(b), the optimized D-H correlation 
weight $f(r)$ for $r=3$ is magnified near the critical points. 
Both $\Psi_{\rm DH}^{\rm exp}$ and $\Psi_{\rm DH}^{\rm pwr}$ exhibit 
critical behavior at the same $U_{\rm c}/t$ as for $\lambda$; 
Jumps in $f(3)$ exist also for $\Psi_{\rm DH}^{\rm opt}$ with $L\ge 20$. 
\par

\begin{figure}[hob]
\vspace{-0.2cm}
\begin{center}
\includegraphics[width=8.5cm,clip]{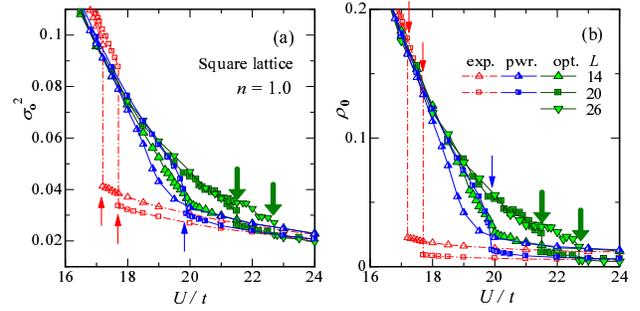} 
\end{center} 
\vskip -5mm
\caption{(Color online) 
(a) Variance of onsite density correlation function (or $2\tilde d$)
and (b) condensate fraction, both calculated using wave functions 
with three kinds of long-range D-H factors, $\Psi_{\rm DH}^{\rm exp}$, 
$\Psi_{\rm DH}^{\rm pwr}$ and $\Psi_{\rm DH}^{\rm opt}$ for a few 
values of $L$. 
The arrows shows the Mott critical points. 
}
\label{fig:cor-cor0-cc}
\end{figure}
%
Next, in Fig.~\ref{fig:cor-cor0-cc}, we show two quantities 
characterizing Mott transitions, $\sigma_{\bf o}$ [eq.~(\ref{eq:N00})] 
or $2\tilde d$ [eq.~(\ref{eq:N00uf})], and $\rho_{\bf 0}$ 
[eq.~(\ref{eq:rho_0})].  
Again, both exhibit anomalous behavior at $U_{\rm c}/t$ determined 
above by the anomalies of variational parameters. 
The behavior is similar to that of QWF (Figs.~\ref{fig:DvsUm} and 
\ref{fig:cor0-cc}). 
Thus, a first-order Mott transition can be described by a wide class 
of D-H binding wave functions if sufficiently large systems 
are considered. 
\par 

Details of the critical behavior as well as the value of Mott critical 
point are different for different wave functions in some degree. 
As clear cusps appear in $E^{\rm exp}$ (Fig.~\ref{fig:EvsUm-cc}), 
$\Psi_{\rm DH}^{\rm exp}$ exhibits sharp first-order critical 
behavior even for a small system of $L=14$. 
Near $U_{\rm c}$, two energy minima are distinguished, and a hysteresis 
is confirmed in the $E/t$-$U/t$ plane. 
A hysteresis is also observed for $\Psi_{\rm DH}^{\rm opt}$ 
widely near $U_{\rm c}/t$, and the energy difference between the 
two phases is very small. 
Thus, the control of optimization process becomes considerably 
difficult in the critical regime, so that the estimation of $U_{\rm c}/t$ 
for $\Psi_{\rm DH}^{\rm opt}$ is less accurate than for the others. 
This is probably caused by the redundancy of parameters. 
In comparison, $\Psi_{\rm DH}^{\rm pwr}$ behaves mildly near 
$U_{\rm c}$ and we have not detect a manifest hysteresis even 
for $L=20$. 
\par 

\begin{table}
\begin{center}
\caption{
The Mott critical values $U_{\rm c}/t$ are compared among the wave 
functions treated in this paper for some system sizes. 
Sections with hyphens indicate the cases in which clear behavior 
of first-order transition is not observed. 
For empty sections, we have not performed calculations in this study. 
The last low shows the result, when the repulsive Jastrow factor 
is included (see \S\ref{sec:LRD-D}). 
In the lower lines, the values obtained in other studies are entered. 
}
\label{table:Uc}
\begin{tabular}{l|l|l|l|l|l|l}
\hline
\multicolumn{1}{c|}{$\Psi$ (or method)} &
\multicolumn{6}{c}{$U_{\rm c}/t$\quad ($n=1$, square lattice)} \\
\hline\hline
\multicolumn{1}{c|}{\qquad\qquad$L\rightarrow$} &
\multicolumn{1}{c|}{10} &
\multicolumn{1}{c|}{14} & 
\multicolumn{1}{c|}{20} &
\multicolumn{1}{c|}{26} & 
\multicolumn{1}{c|}{32} &
\multicolumn{1}{c}{40} \\ 
\hline
QWF    \quad\ ($\Psi_Q$)
       & \multicolumn{1}{c|}{-} & \multicolumn{1}{c|}{-} &
                         20.5 & 21.5 & 22.5 & 24.1 \\
optim.\quad\ ($\Psi_{\rm DH}^{\rm opt}$) 
       & \multicolumn{1}{c|}{-} & \multicolumn{1}{c|}{-} & 
                   $\sim 21.3$ & $\sim 22.7$ &      &      \\
exp.\quad\quad ($\Psi_{\rm DH}^{\rm exp}$) 
       & \multicolumn{1}{c|}{-} & 17.2 & 17.65 &  &  &  \\
power\quad\ ($\Psi_{\rm DH}^{\rm pwr}$)
       & \multicolumn{1}{c|}{-} & \multicolumn{1}{c|}{-} & 
                         19.85 &     &       &     \\
\hline
repulsive\ \ ($\Psi_{\rm J}$) 
       & \multicolumn{1}{c|}{-} & \multicolumn{1}{c|}{-} & 
                         19.85 &     &       &     \\
\hline\hline
\multicolumn{1}{l|}{GWF\ ($U_{\rm BR}/t)$} &
\multicolumn{6}{c}{23.31$\cdots$\ (refs.~\citen{BHM-GWF1} and 
\citen{BHM-GWF2})} \\
\hline
\multicolumn{1}{l|}{VMC (Jastrow)} &
\multicolumn{6}{c}{$\sim 20.6$\ (ref.~\citen{Capello})} \\
\hline
\multicolumn{1}{l|}{Strong coupling} &
\multicolumn{6}{c}{16.7\ (ref.~\citen{Monien})} \\
\hline
\multicolumn{1}{l|}{Recent QMC} &
\multicolumn{6}{c}{16.25\ (ref.~\citen{QMC-Wessel}),\quad 
16.7\ (ref.~\citen{QMC-Sansone})} \\
\hline
\end{tabular}
\vskip -5mm
\end{center}
\end{table}
%
Finally, we discuss the value of the Mott critical point. 
In Table~\ref{table:Uc}, the critical values determined by the D-H 
binding wave functions in this work are listed with those of 
other studies. 
Among the three long-range $\Psi_{\rm DH}$ treated here, 
$U_{\rm c}/t$ is mutually different to some extent. 
On the basis of $E/t$ in Fig.~\ref{fig:EvsUm-cc}, this difference 
is considered to stem from the propriety of $\Psi_{\rm DH}$ in 
describing the superfluid state near $U_{\rm c}/t$. 
Because $E/t$ of the three $\Psi_{\rm DH}$'s is similar to one another 
for $U>U_{\rm c}$, a lower $E/t$ or a better $\Psi_{\rm DH}$ 
in the conductive side of $U_{\rm c}$ yields a larger $U_{\rm c}/t$. 
Furthermore, the critical value $U_{\rm c}/t$, estimated by any 
$\Psi_{\rm DH}$ including QWF (\S\ref{sec:Mott-QWF}), steadily 
increases as $L$ increases. 
This is mainly because the system-size dependence of $E/t$ is 
considerably large in the insulating side of $U_{\rm c}/t$, but small 
in the conductive side, as in Figs.~\ref{fig:EvsUm-c} and 
\ref{fig:EvsUm-cc}. 
These aspects of the critical value are contrary to reliable estimates 
by QMC and a strong-coupling expansion, $U_{\rm c}/t=16$-17. 
To obtain a more accurate critical value, some factor overlooked 
in the present class of wave functions may be taken into account. 
As a possibility, we take up D-D repulsive correlations 
in \S\ref{sec:LRD-D}. 
\par

\subsection{Effect of repulsive Jastrow factors\label{sec:LRD-D}}
So far, we have disregarded the effect of intersite repulsive factors, 
because it is known long-range repulsive factors reduce the energy 
only slightly for fermions at half filling.\cite{YS,Miyagawa} 
Here, we check that the Bose Hubbard model also has this property, 
and a repulsive factor is insufficient to improve the system-size 
dependence of $U_{\rm c}/t$. 
For simplicity, we employ the power-law-decaying type both for 
attractive [eqs.~(\ref{eq:LRDH}) and (\ref{eq:f(r)}b)] and repulsive 
[eqs.~(\ref{eq:Jast}) and (\ref{eq:Jastrow})] correlation factors 
mentioned in \S\ref{sec:formalism}.
\par

\begin{table}
\begin{center}
\caption{
Comparison of total energy per site $E/t$ near $U_{\rm c}/t$ 
among GWF and wave functions with power-law decaying type D-H 
and D-D correlation factors. 
The digits in the brackets indicate the ratios with respect to 
the values of GWF+D-H. 
}
\label{table:D-D}
\begin{tabular}{c|c|c|c}
\hline
\multicolumn{1}{c|}{$U/t$} &
\multicolumn{1}{c|}{GWF} & 
\multicolumn{1}{c|}{GWF+D-H} & 
\multicolumn{1}{c}{GWF+D-H+D-D} \\ 
\hline\hline
\multicolumn{4}{l}{$L=14$} \\
\hline
19 & -0.11712 (0.301) & -0.38945 & -0.39156 (1.005) \\
20 & -0.07435 (0.200) & -0.37103 & -0.37561 (1.012) \\
21 & -0.04314 (0.121) & -0.35523 & -0.35680 (1.004) \\
\hline
\multicolumn{4}{l}{$L=20$\quad ($U_{\rm c}/t=19.85$)} \\
\hline
19 & -0.11042 (0.293) & -0.37742 & -0.37858 (1.003) \\
20 & -0.06750 (0.191) & -0.35410 & -0.35405 (1.000) \\
21 & -0.03612 (0.106) & -0.34034 & -0.34138 (1.003) \\
\hline
\end{tabular}
\vskip -5mm
\end{center}
\end{table}
%
First, we argue improvement in energy by D-H attractive and D-D 
repulsive factors over GWF. 
Table~\ref{table:D-D} lists the numerical values of $E/t$ 
for the three wave functions and two system sizes. 
As $U$ approaches $U_{\rm BR}$ $(=23.31t)$, $E^{\rm GWF}$ steadily 
increases toward zero. 
By introducing the D-H binding factor, $E$ is significantly 
improved on $E^{\rm GWF}$ (70-90\%), and comes to increase slowly 
as $U/t$ increases, like the case of QWF in Fig.~\ref{fig:EvsU-mu}(a). 
On the other hand, when we add the D-D and H-H repulsive correlation
to $\Psi_{\rm DH}$, the improvement of $E/t$ on $\Psi_{\rm DH}$ 
is very slight (mostly less than 1\%) in both conductive and insulating 
phases, and comparable to the statistical errors, which are large 
for $\Psi_{\rm J}$.\cite{note-J} 
\par

\begin{figure}[hob]
\vspace{-0.2cm}
\begin{center}
\includegraphics[width=8.5cm,clip]{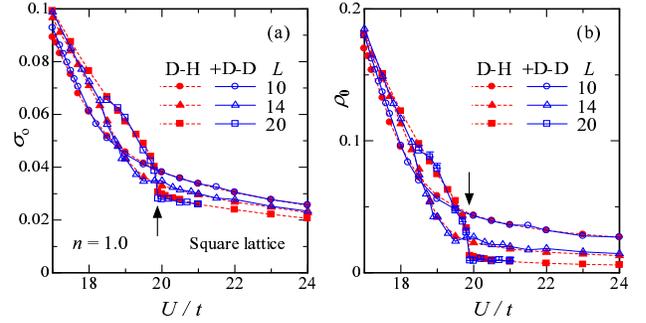} 
\end{center} 
\vskip -5mm
\caption{(Color online) 
(a) The variance of onsite density correlation function (or $2\tilde d$)
and (b) condensate fraction, are compared between 
$\Psi_{\rm DH}^{\rm pwr}$ (D-H) and $\Psi_{\rm J}^{\rm pwr}$ (+D-D) 
for three values of $L$. 
The arrows indicate $U_{\rm c}/t$ for $L=20$. 
}
\label{fig:r-cor-cor0-cc}
\end{figure}
%
Although the energy reduction is slight, we should confirm whether 
the D-D repulsive factor ${\cal P}_J$ improves physical quantities 
relevant to the Mott transition. 
Since ${\cal P}_J$ tends to lengthen the inter-doublon (D-D) distance 
in the insulating side (not shown), we have expected that  
$\Psi_{\rm J}^{\rm pwr}$ lowers $U_{\rm c}/t$. 
In Figs.~\ref{fig:r-cor-cor0-cc}(a) and Figs.~\ref{fig:r-cor-cor0-cc}(b), 
we compare $\sigma_{\rm o}$ and $\rho_{\bf 0}$, respectively, between 
$\Psi_{\rm DH}^{\rm pwr}$ and $\Psi_{\rm J}^{\rm pwr}$. 
Against our anticipation, the differences in both quantities are 
negligibly small, and a meaningful shift of $U_{\rm c}/t$ cannot be 
observed (Table~\ref{table:Uc}). 
\par

Here, we have focused on the power-law decaying case. 
For fermions, it is found that various repulsive factors somewhat lower 
the values of $U_{\rm c}/t$, but the situation in system-size dependence 
does not change.\cite{Miyagawa} 
Hence, we conclude that the effect of repulsive Jastrow factor is 
not significant, as far as the Mott transition is concerned. 
We will take no account of this factor in the following. 
\par

\section{Renewed Picture of Mott Transition\label{sec:picture}}
In a previous paper for electrons,\cite{YOT} we affirmed a simple 
mechanism of Mott transitions, in which the binding of a doublon 
(plus charge carrier) to a holon (minus charge carrier) 
in the insulating regime and the release from binding in the conductive
regime are the essence of the Mott transition. 
In this section, we extend this picture to comprehend a case of 
completely D-H binding. 
\par 

\begin{figure}[hob]
\vspace{-0.2cm}
\begin{center}
\includegraphics[width=8.0cm,clip]{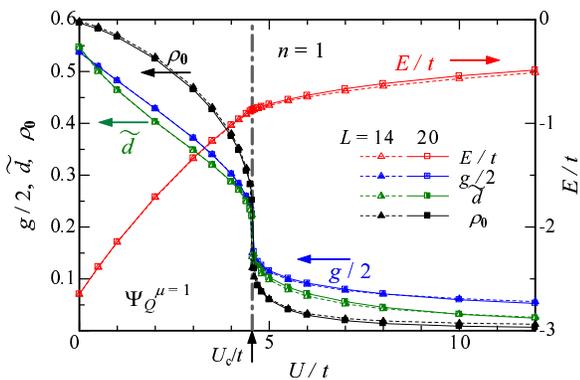} 
\end{center} 
\vskip -5mm
\caption{(Color online) 
Four quantities calculated with the completely D-H bound state
$\Psi_Q^{\mu=1}$ are plotted as a function of $U/t$; namely, 
total energy per site $E/t$ (right axis), half of the optimized 
Gutzwiller parameter $g$, multiplon density $\tilde d$ and condensate 
fraction $\rho_{\bf 0}$. 
The Mott critical point $U_{\rm c}/t$ is indicated by a shadowed 
dash-dotted line. 
The system-size dependence is negligible. 
}
\label{fig:E-dvsU}
\end{figure}

First, we briefly mention the properties of the completely D-H 
bound state in NN sites, $\Psi_Q^{\mu=1}$, namely QWF with 
$\mu(=1)$ and $\mu'(=0)$ in eq.~(\ref{eq:pq-SQL}). 
Because, in $\Psi_Q^{\mu=1}$, a doublon must be accompanied by 
at least one holon in its four NN sites, we tend to regard it 
as insulating for any value of $U/t$. 
In fact, however, $\Psi_Q^{\mu=1}$ is a superfluid state for small 
$U/t$, and exhibit a Mott transition at $U/t=4.55$. 
A sign of a Mott transition can be recognized in the cusp behavior 
of $E/t$ at $U/t\sim 4.55$ in Fig.~\ref{fig:EvsU-mu}(a), and especially 
in sudden drop of the condensate fraction in Fig.~\ref{fig:cor0-cc}(a). 
In Fig.~\ref{fig:E-dvsU}, we show the optimized value of $g$ and 
multiplon density $\tilde d$, besides the above two quantities.  
The discontinuous decreases of the order parameter $\tilde d$, and 
the sudden vanishing of $\rho_{\bf 0}$ corroborate the Mott 
transition at $U_{\rm c}/t=4.55$. 
\par

%
This Mott transition in $\Psi_Q^{\mu=1}$, especially in the conductive 
side, cannot be understood by a simple release from the D-H binding 
and mutually independent movement of the two kinds of carriers, because 
a doublon is in contact with a holon even in the conductive regime. 
Instead of this simple point of view, we propose an improved picture 
extensively applicable to Mott transitions. 
To this end, it is convenient to introduce two relevant length scales, 
D-H binding length $\xi_{\rm dh}$, and minimum D-D (H-H) distance 
$\xi_{\rm dd}$, which are loosely defined as follows: 
The distances of most of the nearest D-H pairs are smaller than 
$\xi_{\rm dh}$; in other words, a doublon is seldom distant 
from a holon beyond $\xi_{\rm dh}$. 
$\xi_{\rm dd}$ is a D-D (H-H) exclusion distance, namely, the 
inter-doublon (inter-holon) distances within which two doublons 
(or holons) are mutually almost inaccessible. 
In general, $\xi_{\rm dh}$ as well as $\xi_{\rm dd}$ depends on $U/t$. 
\par

\begin{figure}[!t] 
\vspace{1.0cm} 
\begin{center}
\includegraphics[width=8.5cm,clip]{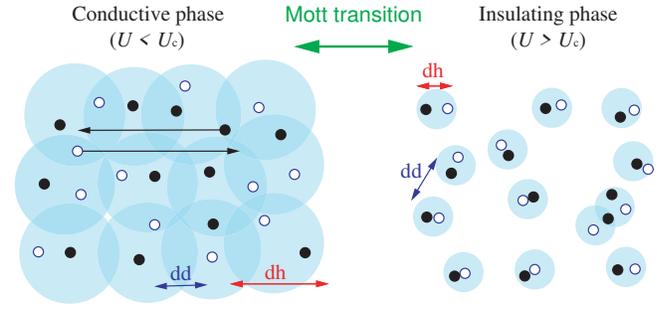} 
\end{center} 
\vskip -15mm
\caption{(Color online) 
Schematic figure of improved picture of Mott transitions. 
Small solid and open circles denote doublons and holons, respectively. 
The D-D distance is closely related to $\xi_{\rm dd}$, shown by ``dd''. 
Each pale circle indicates the domain of a D-H pair, whose diameter 
roughly corresponds to $\xi_{\rm dh}$, shown by ``dh". 
Typical configurations are drawn for the two phases. 
In each pair domain, the existence of at least one doublon and one holon 
is required, but excess doublons and holons can move around independently
slipping out of the original domain, as indicated by long arrows 
for the conductive phase. 
} 
\label{fig:pictureai} 
\end{figure} 
%
Then, we postulate that an attractive correlation factor ${\cal P}_Q$ 
yields D-H pairs of a binding length $\xi_{\rm dh}$ according to $U/t$. 
An improved picture of Mott transition is schematically represented 
in Fig.~\ref{fig:pictureai}. 
In the insulating phase, the relation $\xi_{\rm dh}<\xi_{\rm dd}$ holds, 
indicating that the domains of D-H pairs do not usually overlap, 
at least, not in sequence. 
As a result, most D-H pairs are isolated and a doublon and a holon 
are confined within $\xi_{\rm dh}$, resulting in only local density 
fluctuation. 
To this point, the picture is basically identical with one previously 
proposed. 
In the conductive phase ($U<U_{\rm c}$), $\xi_{\rm dh}$ becomes longer 
than $\xi_{\rm dd}$, indicating the domains of D-H pairs overlap 
with one another. 
Then, a doublon in a D-H pair can exchange a partner holon with a holon 
in an adjacent D-H pair. 
Consequently, a doublon and a holon can move independently as carriers 
by exchanging the partner, as shown in long arrows 
in Fig.~\ref{fig:pictureai}. 
It follows that, as the value of $U/t$ is varied, a Mott transition 
takes place when $\xi_{\rm dh}$ becomes equivalent to $\xi_{\rm dd}$, 
which (correctly $\ell_{\rm dd}$) is roughly $1/\sqrt{\tilde d}$, 
and is expected to be a monotonically increasing function of $U/t$. 
\par

To justify the above picture, we have to appropriately estimate 
$\xi_{\rm dh}$ and $\xi_{\rm dd}$, and confirm that the Mott transitions 
take place when these two length scales intersect each other 
at $U_{\rm c}/t$. 
After checking various cases, we have found that the following 
formulae work well and are physically natural to the above point of 
view,
\begin{equation}
\xi_{\rm dh}=\ell_{\rm dh}+\sigma_{\rm dh}, 
\label{eq:dh}
\end{equation}
\begin{equation}
\xi_{\rm dd}=\ell_{\rm dd}-\sigma_{\rm dd}, 
\label{eq:dd}
\end{equation}
where $\ell_{\rm dh}$ ($\ell_{\rm dd})$ denotes the average of 
the nearest D-to-H and H-to-D (D-to-D and H-to-H) distance, 
and $\sigma_\Lambda$ the standard deviation of $\ell_\Lambda$, 
with $\Lambda$ being an index of ``dh" and ``dd": 
\begin{equation}
\sigma_\Lambda
  =\sqrt{\frac{1}{M}\sum_{m=1}^M\left(\ell_\Lambda^m-\ell_\Lambda\right)^2}
.
\label{eq:sd}
\end{equation}
Here, $m$ runs over all the doublons and holons in all the measured 
samples, $M$ indicates the total number of doublons and holons 
in all the measured samples, and $\ell_\Lambda^m$ indicates the nearest 
$\Lambda$ distance for the $m$-th doublon (or holon).
The addition and subtraction of $\sigma$ in eqs.~(\ref{eq:dh}) and 
(\ref{eq:dd}) represent a $U/t$-dependent ``softness" of the D-H binding 
and D-D repulsive correlations, respectively. 
\par

\begin{figure}[!t] 
\vspace{0.5cm} 
\begin{center}
\includegraphics[width=8.5cm,clip]{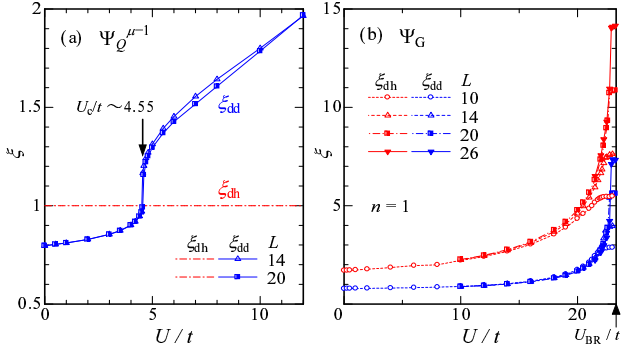} 
\end{center} 
\vskip -3mm
\caption{(Color online) 
The D-H binding length $\xi_{\rm dh}$ and minimum D-D distance 
$\xi_{\rm dd}$ defined by eqs.~(\ref{eq:dh}) and (\ref{eq:dd}) are 
plotted versus $U/t$ for 
(a) the completely D-H bound state, and 
(b) the Gutzwiller wave function, 
for some values of $L$. 
} 
\label{fig:distan-mu-g} 
\end{figure}
\begin{figure}[!t] 
\vspace{0.5cm} 
\begin{center}
\includegraphics[width=7.0cm,clip]{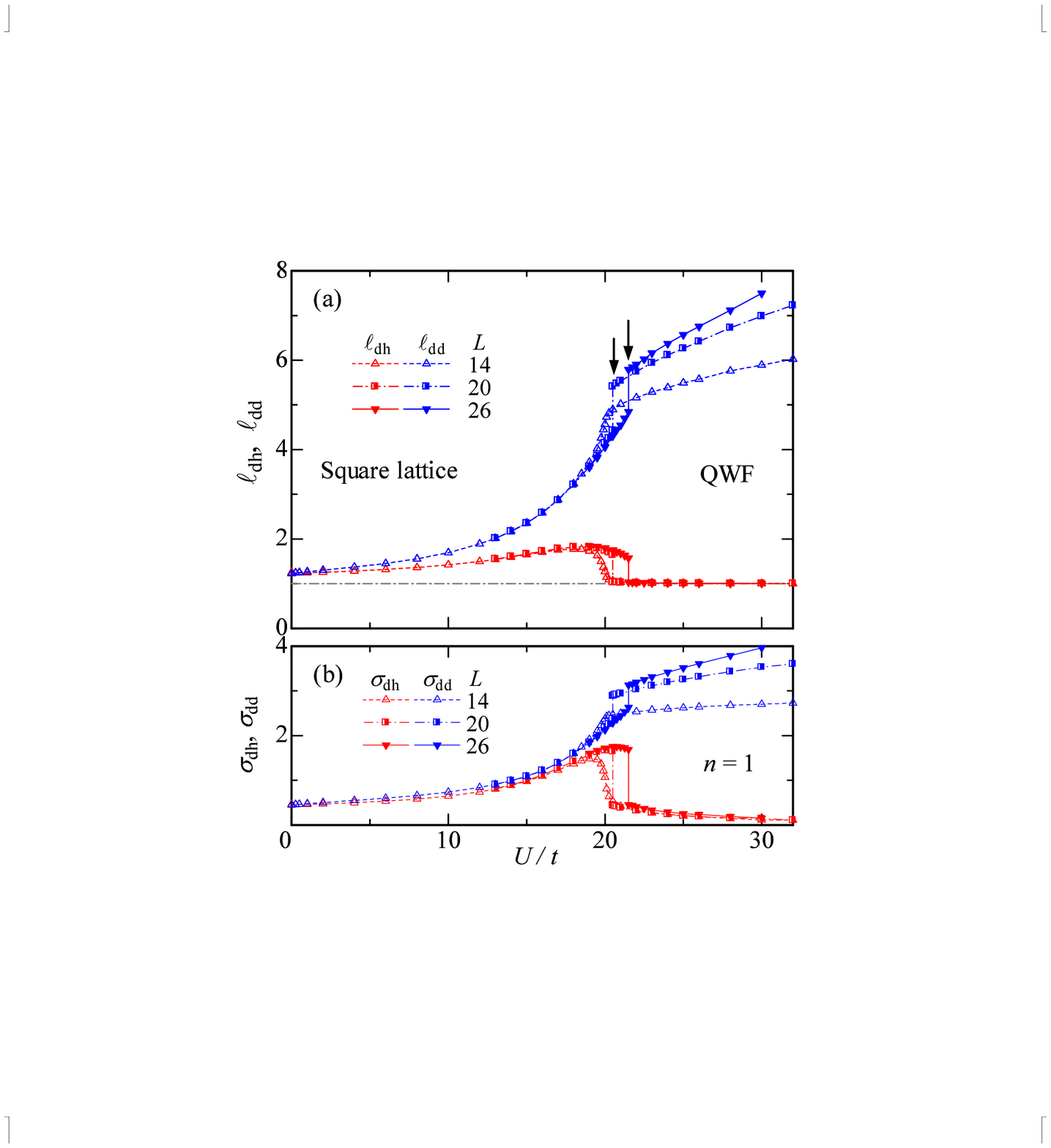} 
\end{center} 
\vskip -3mm
\caption{(Color online) 
(a) The average nearest D-H and D-D distances and (b) their 
standard deviations are plotted for the short-range D-H binding 
wave function QWF as a function of $U/t$. 
The Mott critical values for $L=20$ and 26 are indicated by arrows 
in (a). 
The scales of coordinates in (a) and (b) are common. 
} 
\label{fig:distan-l-sd} 
\end{figure}

 
%
First, we discuss a special case of the tightly D-H bound 
state $\Psi_Q^{\mu=1}$, shown in Fig.~\ref{fig:distan-mu-g}(a). 
Because $\ell_{\rm dh}$ is restricted to 1 in $\Psi_Q^{\mu=1}$, 
$\sigma_{\rm dh}=0$ and $\xi_{\rm dh}=1$ hold, irrespective of 
the value of $U/t$. 
On the other hand, because the doublon density decreases 
as $U/t$ increases, both $\ell_{\rm dd}$ and $\sigma_{\rm dd}$ 
monotonically increase qualitatively in a manner similar to 
those of QWF shown in Fig.~\ref{fig:distan-l-sd}, and 
consequently, $\xi_{\rm dd}$ also monotonically increases 
with a discontinuity at $U_{\rm c}/t$. 
As shown in Fig.~\ref{fig:distan-mu-g}(a), $\xi_{\rm dd}$ 
intersects $\xi_{\rm dh}$ in this discontinuity; this special 
case is consistent with the above picture. 
\par 

%
\begin{figure}[!t] 
\vspace{0.5cm} 
\begin{center}
\includegraphics[width=8.5cm,clip]{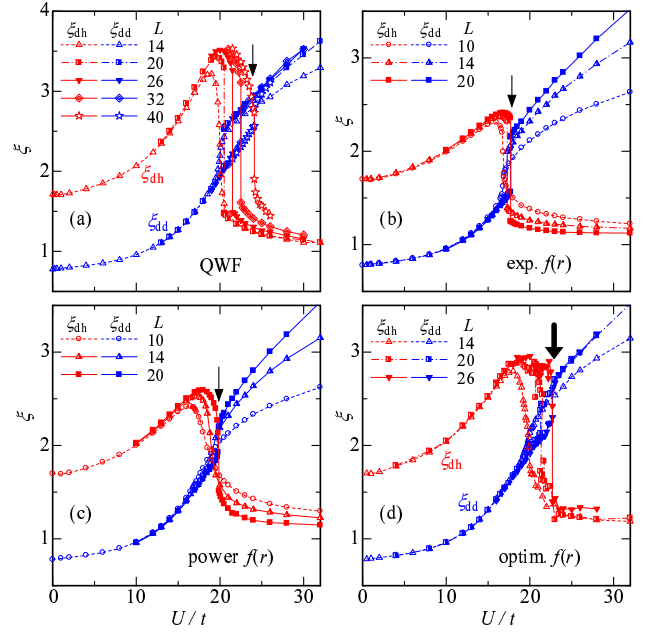} 
\end{center} 
\vskip -3mm
\caption{(Color online) 
The D-H binding length $\xi_{\rm dh}$ and minimum D-D distance 
$\xi_{\rm dd}$ defined by eqs.~(\ref{eq:dh}) and (\ref{eq:dd}) are 
plotted versus $U/t$ for 
(a) the short-range D-H binding state QWF, 
(b) the exponentially-decaying D-H binding state $\Psi_{\rm DH}^{\rm exp}$, 
(c) the power-law-decaying D-H binding state $\Psi_{\rm DH}^{\rm pwr}$,
(d) the optimized D-H binding state $\Psi_{\rm DH}^{\rm opt}$,
for some values of $L$. 
An arrow in each panel denotes the Mott critical point for the largest 
system size in each state.
} 
\label{fig:distan-sd-cc} 
\end{figure} 
%
Next, we consider more ordinary D-H binding wave functions: 
QWF, $\Psi_{\rm DH}^{\rm exp}$, $\Psi_{\rm DH}^{\rm pwr}$ and 
$\Psi_{\rm DH}^{\rm opt}$. 
In Fig.~\ref{fig:distan-l-sd}, we show the average nearest D-H and 
D-D distances and their standard deviations for QWF. 
The behavior of $\ell$'s and of $\sigma$'s for the three long-range 
D-H binding wave functions is basically the same. 
As $U/t$ increases, $\ell_{\rm dd}$ rapidly increases, 
exhibits a discontinuity at $U_{\rm c}/t$, and further increases 
for $U>U_{\rm c}$ with a considerable system-size dependence. 
Meanwhile, $\ell_{\rm dh}$ at first gradually increases, and has 
a maximum in the vicinity of $U_{\rm c}/t$, then decreases 
discontinuously at $U_{\rm c}/t$, and converges to $\ell_{\rm dh}=1$ 
for $U>U_{\rm c}$. 
The magnitude of $\sigma_{\rm dh}$ becomes large near $U_{\rm c}/t$ 
in the conductive side, but tends to vanish for $U>U_{\rm c}$. 
Substituting such $\ell$ and $\sigma$ for eqs.~(\ref{eq:dh}) and 
(\ref{eq:dd}), we obtain $\xi_{\rm dh}$ and $\xi_{\rm dd}$ 
for each of the four D-H binding states, and plot them 
in Fig.\ref{fig:distan-sd-cc}. 
As expected from Fig.~\ref{fig:distan-l-sd}, $\xi_{\rm dd}$ is 
a monotonically increasing function of $U/t$ with a discontinuity 
at $U_{\rm c}$, whereas $\xi_{\rm dh}$ has a maximum just below 
$U_{\rm c}$ and becomes a decreasing function of $U/t$ 
with a discontinuous drop at $U_{\rm c}$. 
Consequently, the magnitude of the two length scales is reversed 
suddenly at $U_{\rm c}$, and the relation $\xi_{\rm dd}>\xi_{\rm dh}$ 
holds for $U>U_{\rm c}$ for each $L$ of every wave function. 
Thus, we conclude that a picture explained in Fig.~\ref{fig:pictureai}
seems appropriate to the D-H binding mechanism of Mott transitions. 
\par 

Incidentally, the above point of view is out of focus for a non-D-H 
binding mechanism like Brinkman-Rice-type transitions. 
As depicted in Fig.~\ref{fig:distan-mu-g}(b), for GWF, $\xi_{\rm dh}$ 
is always longer than $\xi_{\rm dd}$, and does not show a tendency 
to decrease, up to the Brinkman-Rice point. 
The value of $\xi$ at $U_{\rm BR}/t$ is proportional to $L$, which 
means that $\Psi_{\rm G}$ entirely lacks intersite correlations. 
\par 

\section{Summary\label{sec:sum}}
In this paper, we have studied the spinless Bose Hubbard models 
at unit filling on the square and triangular lattices, using 
a variational Monte Carlo scheme. 
Our primary aim is to grasp fundamentals of the Mott transition 
without influence of the spin degree of freedom. 
In the trial wave functions, we allow for various types of doublon-holon 
attractive factors and a doublon-doublon (holon-holon) repulsive factor 
in addition to the onsite repulsive (Gutzwiller) factor. 
We itemize the main results below.
\par

(1) Because multiply occupied sites with more than two particles 
almost vanish for $U\gsim U_{\rm c}/2$ for all the wave functions 
we check, we can consider the Mott physics near $U_{\rm c}/t$ ($\sim 20$
for the square lattice) only with a doubly occupied site (doublon, D), 
an empty site (holon, H) and a singly occupied site, like fermionic 
cases.
\par

(2) Wave functions with appropriate D-H attractive correlation 
factors exhibit first-order Mott transitions. 
Unlike the Brinkman-Rice-type transition arising in the Gutzwiller 
wave function, these transitions have density fluctuation in the Mott 
insulator side. 
By density fluctuation, anisotropy is introduced into the momentum 
distribution function, which is consistent with absorption images 
observed in cold atom experiments.\cite{Greiner} 
The total energy for $U>U_{\rm c}$ coincides well with 
the results of the strong coupling expansion ($E/t\propto -t/U$). 
\par

(3) In the conductive (superfluid) state, the optimized value of D-H 
attractive correlation weight $f(r)$ [eq.~(\ref{eq:LRDH})], 
with $r$ being the interparticle 
distance, rapidly decreases with $r$ for $r\lsim 3$, and is almost 
constant for larger $r$. 
Thereby, we found it reasonable that the wave function with short-range 
D-H factors (QWF), in which $f(r)$ for $r\ge 3$ is constant, 
gives unexpectedly good results.  
On the other hand, in the Mott insulating state, $f(r)$ almost vanishes 
for $r\ge 3$, suggesting that a doublon and a holon confine 
each other in near-neighbor ($r\le 2$) sites as a D-H pair. 
This will be confirmed by recently developed quantum gas microscope 
experiments.\cite{Bakr,Sherson} 
\par 

(4) The critical behavior in the triangular lattice is more 
continuous-transition-like than in the square lattice. 
A similar phenomenon emerges in paramagnetic states of electronic 
systems,\cite{YOT} and was ascribed to the frustration of spins. 
However, it is probable that its origin is closely related to 
the connectivity of lattices. 
\par 

(5) We have improved the D-H binding picture of Mott transitions, 
by introducing two characteristic length scales, D-H binding length 
$\xi_{\rm dh}$, which broadly represents the size of a D-H pair, 
and minimum D-D distance $\xi_{\rm dd}$. 
We appropriately determined $\xi_{\rm dh}$ and $\xi_{\rm dd}$. 
In the conductive state ($\xi_{\rm dh}>\xi_{\rm dd}$), because 
the domains of D-H pairs mutually overlap, density carriers (D and H) 
can move independently of the partners of the D-H pairs released 
from the binding. 
In the insulating state ($\xi_{\rm dh}<\xi_{\rm dd}$), because most 
domains of D-H pairs are detached from one another, density fluctuation 
is localized within the domain of $\xi_{\rm dh}$. 
The Mott transition takes place, when the relation 
$\xi_{\rm dh}=\xi_{\rm dd}$ is satisfied. 
\par

(6) By adding a D-D (and H-H) repulsive factor, the variational energy 
is improved slightly, especially in the insulating regime. 
However, we have not recognized qualitative influences thereof
on the Mott transition. 
\par

Since the properties of the Mott transition studied in this paper 
basically coincide with those for the electron systems,\cite{YOT} 
the renewed picture of Mott transitions given in \S\ref{sec:picture} 
can be applied to electron systems.\cite{Miyagawa}
An important remaining problem is to determine the Mott critical 
value $U_{\rm c}/t$ more accurately. 
This problem is closely connected with the great system-size 
dependence of the present trial wave functions around $U_{\rm c}/t$. 
To remedy it, we may need more exquisite size-dependent correlation 
factors in the Mott critical regime.\cite{Tahara}. 
\par

\begin{acknowledgments}
We would like to thank Hiroki Tsuchiura, Yuta Toga (Tohoku University), 
Makoto Yamashita (NTT) and Kenji Kobayashi (Chiba Institute of Technology) 
for useful discussions and information. 
This work is partly supported by Grant-in-Aids from the Ministry of 
Education, Culture, Sports, Science and Technology. 
\end{acknowledgments}

\appendix
\section{Variational Monte Carlo Method} 
We briefly note the setup and condition of VMC calculations 
carried out in this paper. 
\par

Because we need to optimize variational parameters up to the maximum 
number $L$, we use a correlated-measurement or optimization-VMC 
technique.\cite{Umrigar} 
In the non-linear minimization process of energy expectation values, 
we adopt a quasi-Newton method, in which gradient vectors are effectively 
calculated, especially for bosons, by recently proposed 
formulae,\cite{Umrigar-Filippi} and Hessian matrices are 
approximately given by Broyden-Flecher-Goldfarb-Shanno 
formula,\cite{Fletcher} the use of which does not affect the accuracy 
of optimization itself. 
In coding, we refer to an algorithm offered by Ibaraki and 
Fukushima.\cite{Ibaraki} 
For wave functions with a few parameters, we use a simple 
linear optimization together. 
\par 

In both algorithms, parameters as well as energy converges typically 
after first several rounds of iteration with different fixed sample 
sets; in each set we generate $2.4$-$2.5\times10^5$ particle 
configurations with Metropolis algorithm. 
After this convergence, we continue excess rounds (15-90 times) 
of iteration in the optimization process with successively renewed 
configuration sets. 
We determine the optimized values by averaging the data obtained 
in the excess rounds; in averaging, we exclude scattered data 
beyond the range of twice the standard deviation. 
Thus, the optimal value is substantially an average of more than 
several million samples. 
The variational energy and significant parameters [$g$ and $f(2)$ etc.] 
are determined with sufficient accuracy in most cases, 
but accurate determination of insignificant parameters [$f(r)$ with 
$r>7$ etc.] is difficult, because $E$ depends on them only very 
slightly, in other words, particle configurations determining them 
appear extremely rarely. 
Anyway, such parameters have little influence on $E$ and other 
quantities. 
Physical quantities are calculated with $2.4$-$2.5\times10^5$ 
renewed configurations generated by the optimized parameter sets. 
\par

Since in Mott critical regimes, the global minimum becomes more 
competitive with other minima as $L$ increases, accurate energy 
minimization sometimes becomes not easy, especially for the triangular 
lattice with QWF and for $\Psi_{\rm DH}^{\rm opt}$ and $\Psi_{\rm J}$. 
This is the cause of scattered data points near $U_{\rm c}/t$ 
in some figures. 
\par



\begin{thebibliography}{99}


\bibitem{Fisher} For instance, M.~P.~A.~Fisher, P.~B.~Weichman, 
G.~Grinstein and D.~S.~Fisher: 
\journal{\PRB}{40}{546}{1989}. 

\bibitem{Greiner} M.~Greiner, O.~Mandel, T.~Esslinger, 
T.~W.~H\"ansch and I.~Bloch: 
\journal{Nature}{415}{39}{2002}. 

\bibitem{1D} T.~St\"oferle, H.~Moritz, C.~Schori, M.~K\"ohl and 
T.~Esslinger: \journal{\PRL}{92}{130403}{2004}.

\bibitem{Kohl} M.~K\"ohl, H.~Moritz, T.~St\"oferle, C.~Schori and 
T.~Esslinger: 
\journal{\JLTP}{138}{635}{2005}. 

\bibitem{Spielman} I.~B.~Spielman, W.~D.~Phillips and J.~V.~Porto: 
\journal{\PRL}{98}{080404}{2007}.

\bibitem{Jaksch} D.~Jaksch, C.~Bruder, J.~I.~Cirac, C.~W.~Gardiner 
and P.~Zoller: 
\journal{\PRB}{81}{3108}{1998}.

\bibitem{Bloch-Rev} I.~Bloch, J.~Dalibard and W.~Zwerger: 
\journal{\RMP}{80}{885}{2008}.

\bibitem{Lieb-Wu} For one dimension, E.~H.~Lieb and F.~Y.~Wu: 
\journal{\PRL}{20}{1445}{1968}.

\bibitem{critical} At unit filling, the critical value is estimated 
at $U_{\rm c}/t\sim 3.6$ in one dimension (ref.~\citen{Uc-1d}), 
and at $U_{\rm c}/t\sim 29.3$ (ref.~\citen{Uc-3d1}) or $31.3$
(ref.~\citen{Uc-3d2}) for the simple cubic lattice. 

\bibitem{Uc-1d} T.~D.~K\"uhner and H.~Monien: 
\journal{\PRB}{58}{R14741}{1998}.

\bibitem{Uc-3d1} B.~Capogrosso-Sansone, N.~V.~Prokof'ev and 
B.~V.~Svistunov:
\journal{\PRB}{75}{134302}{2007}. 

\bibitem{Uc-3d2} Y.~Kato, Q.~Zhou, N.~Kawashima and N.~Trivedi: 
\journal{\NP}{4}{617}{2008}.

\bibitem{QMC-Krauth} W.~Krauth and N.~Trivedi: 
\journal{\EPL}{14}{627}{1991}.

\bibitem{QMC-Wessel} S.~Wessel, F.~Alet, M.~Troyer and G.~G.~Batrouni: 
\journal{\PRA}{70}{053615}{2004}. 

\bibitem{QMC-Sansone} B.~Capogrosso-Sansone, S.~G.~S\"oyler, N.~Prokof'ev 
and B.~Svistunov:
\journal{\PRA}{77}{015602}{2008}. 

\bibitem{Monien} N.~Elstner and H.~Monien: 
\journal{\PRB}{59}{12184}{1999}. 


\bibitem{McMillan} W.~L.~McMillan: \journal{\PR}{138}{A442}{1965}.

\bibitem{YS1} H.~Yokoyama and H.~Shiba: \journal{\JPSJ}{56}{1490}{1987}.

\bibitem{Umrigar} C.~J.~Umrigar, K.~G.~Wilson and J.~W.~Wilkins: 
\journal{\PRL}{60}{1719}{1988}. 

\bibitem{Gutz} M.~Gutzwiller: 
\journal{\PRL}{10}{159}{1963}.

\bibitem{BHM-GWF1} W.~Krauth, M.~Caffarel, J.~-P.~Bouchaud: 
\journal{\PRB}{45}{3137}{1992}. 

\bibitem{BHM-GWF2} D.~S.~Rokhsar and B.~G.~Kotliar: 
\journal{\PRB}{44}{10328}{1991}. 

\bibitem{GA} M.~Gutzwiller, \journal{\PR}{137}{A1726}{1965}. 

\bibitem{BR} W.~F.~Brinkman and T.~M.~Rice: 
\journal{\PRB}{2}{4302}{1970}. 

\bibitem{Capello} M.~Capello, F.~Becca, M.~Fabrizio and S.~Sorella:
\journal{\PRL}{99}{056402}{2007}, and 
\journal{\PRB}{77}{144517}{2008}.

\bibitem{Castellani} C.~Castellani, C.~Di Castro, D.~Feinberg and 
J.~Ranninger: \journal{\PRL}{43}{1957}{1979}. 

\bibitem{Kaplan}
T.~A.~Kaplan, P.~Horsch and P.~Fulde: 
\journal{\PRL}{49}{889}{1982}. 

\bibitem{Fazekas} P.~Fazekas and K.~Penc:
\journal{\IJMP}{B1}{1021}{1988}; P.~Fazekas, 
\journal{Physica Scripta T}{29}{125}{1989}. 

\bibitem{YS} H.~Yokoyama and H.~Shiba:
\journal{\JPSJ}{59}{3669}{1990}.

\bibitem{Y-PTP} H.~Yokoyama: \journal{\PTP}{108}{59}{2002}. 

\bibitem{YTOT} H.~Yokoyama, Y.~Tanaka, M.~Ogata and H.~Tsuchiura: 
\journal{\JPSJ}{73}{1119}{2004}. 

\bibitem{YOT} H.~Yokoyama, M.~Ogata and Y.~Tanaka: 
\journal{\JPSJ}{75}{114706}{2006}.

\bibitem{Watanabe} T.~Watanabe, H.~Yokoyama, Y.~Tanaka and J.~Inoue: 
\journal{\JPSJ}{75}{074707}{2006}.

\bibitem{Miyagawa} T.~Miyagawa and H.~Yokoyama: to appear in 
Physica C (2011), and submitted to \JPSJ 

\bibitem{SNS2007} H.~Yokoyama and M.~Ogata: 
\journal{\JPCS}{69}{3356}{2008}.

\bibitem{YMO} H.~Yokoyama, T.~Miyagawa and M.~Ogata: to appear in 
Physica C (2011).

\bibitem{box} For instance, 
T.~P.~Meyrath, F.~Schreck, J.~L.~Hanssen, C.-S.~Chuu and M.~G.~Raizen: 
\journal{\PRA}{71}{041604}{2005}.

\bibitem{errata} In the previous papers, refs.~\citen{YOT}, 
\citen{Watanabe}, and \citen{SNS2007}, we carelessly made a mistake 
in the expression of operators $Q$. 
The correct expression is like in eq.~(\ref{eq:Q}) in this paper. 

\bibitem{notemu'} The parameter $\mu'$ for diagonal neighbors (a part 
of the sites with $r=2$) works irrespective of the particle configuration 
in the NN sites ($r=1$). 
In this point, the effect of $\mu'$ is not identical with the weight 
$f(2)$ discussed in the item (2). 

\bibitem{Harris} For instance, 
A.~B.~Harris and R.~V.~Range: \journal{\PR}{157}{295}{1967}. 

\bibitem{number-Ger} F.~Gerbier, S.F\"olling, A.~Widera, O.~Mandel and 
I.~Bloch:
\journal{\PRL}{96}{090401}{2006}. 

\bibitem{number-Che} P.~Cheinet, S.~Trotzky, M.~Feld, U.~Schnorrberger, 
M.~Moreno-Cardoner, S.~F\"olling and I.~Broch: 
\journal{\PRL}{101}{090404}{2008}. 

\bibitem{Lu} X.~Lu and Y.~Yu: \journal{\PRA}{74}{063615}{2006}.

\bibitem{Sansone} B.~Capogrosso-Sansone, E.~Kozik, N.~Prokof'ev 
and B.~Svistunov: \journal{\PRA}{75}{013619}{2007}. 

\bibitem{Bakr} W.~S.~Bakr, A.~Peng, M.~E.~Tai, R.~Ma, J.~Simon, 
J.~I.~Gillen, S.~F\"olling, L.~Pollet, M.~Greiner: 
\journal{Science}{329}{547}{2010}. 

\bibitem{Sherson} J.~F.~Sherson, C.~Weitenberg, M.~Endres, M.~Cheneau and 
I.~Bloch: \journal{Nature}{467}{68}{2010}.

\bibitem{Kapit} E.~Kapit and E.~Mueller: 
\journal{\PRA}{82}{013644}{2010}.

\bibitem{Mun} J.~Mun, P.~Medley, G.~K.~Campbell, L.~G.~Marcassa, 
D.~E.~Pritchard and W.~Ketterle:
\journal{\PRL}{99}{150604}{2007}.

\bibitem{Spielman2} I.~B.~Spielman, W.~D.~Phillips and J.~V.~Porto: 
\journal{\PRL}{100}{120402}{2008}.

\bibitem{Schroll} C.~Schroll, F. Marquardt and C.~Bruder: 
\journal{\PRA}{70}{053609}{2004}.

\bibitem{noteDH} Precisely speaking, because the symmetry between 
D and H is broken for $n\ne 1$, we should differentiate the correlation 
factors between D to H and H to D. 
However, here we use the identical factor for both, because the 
difference is negligible even quantitatively for small $|1-n|$, 
as we checked in ref.~\citen{YOTKT} for electronic systems. 

\bibitem{YOTKT} H.~Yokoyama, M.~Ogata, Y.~Tanaka, K.~Kobayashi and 
H.~Tsuchiura: in preparation. 

\bibitem{note-J} 
In the repulsive factor $g^D{\cal P}_{\rm J}$, the onsite repulsive 
correlation, mainly controlled by $g$, is affected also by the long-range 
part of the repulsive factor ${\cal P}_{\rm J}$, 
namely the parameter set has redundancy. 
Consequently, the optimized parameter set are sometimes not uniquely 
determined, leading to large statistical errors. 
In this case, the form used in ref.~\citen{YS} will be better. 
One should avoid the redundancy in the trial function.


\bibitem{Tahara} D.~Tahara and M.~Imada: 
\journal{\JPSJ}{77}{093703}{2008}, and 
\journal{\JPSJ}{77}{114701}{2008}. 



\bibitem{Umrigar-Filippi} C.~J.~Umrigar and C.~Filippi: 
\journal{\PRL}{94}{150201}{2005}; 
S.~Sorella: \journal{\PRB}{71}{241103}{2005}. 

\bibitem{Fletcher} For instance, 
R.~Fletcher: {\it Practical Methods of Optimization} 2nd ed., 
(John Wily, Chichester, 1987). 

\bibitem{Ibaraki} T.~Ibaraki and M.~Fukushima: 
{\it FORTRAN77 Optimization Programming}, chap.~6 (Iwanami, Tokyo, 1991), 
[in Japanese].



\end{thebibliography}
\end{document}